\documentclass[12pt]{article}
\usepackage[normalem]{ulem}
\usepackage{amsmath}
\usepackage{amssymb}
\usepackage{color}
\usepackage{cite}
\begin{document}
\begin{center}
{\large\bf  { Discrete and Continuous Symmetry Transformation Operators and Their Algebraic Structures: A $3D$ Field-Theoretic System}}

\vskip 2.0cm

{\sf R. Kumar$^{(a)}$, R. P. Malik$^{(b,c)}$}\\
$^{(a)}${\textit {Department of Physics, Siksha Bhavana,\\
Visva-Bharati, Santiniketan, Bolpur--731235, West Bengal, India}}\\

\vskip 0.2cm
$^{(b)}${\textit {Department of Physics, Institute of Science,\\ Banaras Hindu University,
Varanasi--221005, Uttar Pradesh (U. P.), India}}\\
\vskip 0.2cm

$^{(c)}${\textit{DST Centre for Interdisciplinary Mathematical Sciences,\\
 Institute of Science, Banaras Hindu University, Varanasi--221005, U. P., India}}\\
 \vskip 0.1cm
 
{\small {\sf {e-mails: rohit.kumar@visva-bharati.ac.in; rpmalik1995@gmail.com}}}

\end{center}

\vskip 2cm

\noindent
We discuss the discrete and continuous symmetry transformation operators
 for a three $(2 + 1)$-dimensional $(3D)$ {\it combined} 
field-theoretic system of the free Abelian 1-form and 2-form {\it gauge} theories within the framework of 
Becchi--Rouet--Stora--Tyutin (BRST) formalism and establish their relevance in the context of the
{\it algebraic} structures that are obeyed by the de Rham cohomological operators of differential 
geometry. In fact, our present field-theoretic system respects {\it six} continuous symmetry
transformations and {\it a couple} of very useful discrete duality symmetry transformations.
Out of the above six continuous symmetry transformations, {\it four} are off-shell nilpotent (i.e.
fermionic) in nature and {\it two} are bosonic. The algebraic structures, obeyed by the symmetry 
operators, are reminiscent of the algebra satisfied by the de Rham cohomological
operators. Hence, our present $3D$ field-theoretic system provides a perfect example for
Hodge theory where there is a convergence of ideas from the physical aspects of the BRST
formalism and mathematical ingredients that are associated with the cohomological operators  
of differential geometry at the {\it algebraic} level. One of the highlights of our present investigation is the
appearance of a massless pseudo-scalar field in our theory (on the symmetry grounds {\it alone}) which
carries the {\it negative} kinetic term. Thus, it is one of the possible candidates for the ``phantom"  
fields of the cyclic, bouncing and self-accelerated cosmological models of the Universe.


\newpage


\section{Introduction}
\label{sec1}


It is a well-known fact that, at the initial stages of the developments in theoretical physics, 
the basic concepts and ideas were borrowed from pure mathematics. However, at present, 
this scenario has changed dramatically due to the mathematical sophistication and rigor required in the realm of modern-day theoretical high energy physics (THEP) to explain some of the excellent experimental observations that have been made in the domain of high energy physics. To be specific, the enormous progress made in the theoretical studies of (super)string theories (see, e.g.~\cite{gsw,pol,lus,bbs,ric} and references therein) has brought together, 
in an unprecedented manner, the active investigators in the research areas of THEP and pure mathematics almost on the {\it same} 
intellectual footing. In other words, there has been a notable convergence of ideas from pure mathematics and THEP 
which has been fruitful to {\it both} kinds of active researchers. In our studies on the field-theoretic examples 
for Hodge theory, such a kind of convergence of ideas has been observed where the physical aspects of 
Becchi--Rouet--Stora--Tyutin (BRST) formalism~\cite{brs1,brs2,brs3,brs4} and mathematical ingredients of the de Rham 
cohomological operators of differential geometry (see, e.g.~\cite{egh,mm,gss,van,nis,joel}) 
have been found to blend together in a beautiful fashion. To be precise, we have been able 
to demonstrate that the {\it massless} and St{\"u}ckelberg-modified {\it massive} Abelian $p$-form ($p = 1, 2, 3$)
gauge theories in $D = 2p$ (i.e. $D = 2, 4, 6$) dimensions of spacetime are the tractable field-theoretic
models of  Hodge theory (see, e.g.~\cite{rpm2,rpm3,rpm4,rpm5}  and references therein) within the framework 
of BRST formalism\footnote{It is worthwhile to point  out that such studies have 
{\it also} been shown to be connected to the field-theoretic models of the topological field theories~\cite{tho} 
because we have been able to propose a {\it new} type of BRST-quantized $2D$ topological field theory (TFT) which captures~\cite{rpm1} a few
aspects of the Witten-type TFT~\cite{wit} and some salient features of the Schwarz-type TFT~\cite{sch}.  Hence, 
our studies  have been physically as well as mathematically useful in a subtle manner.} where the discrete and continuous symmetry transformations (and corresponding conserved charges) have 
been able to provide the physical realizations of the de Rham cohomological operators of differential geometry at the 
{\it algebraic}  level. It is pretty obvious that these kinds of field-theoretic models of Hodge theory  are defined 
{\it only} in the {\it even} dimensions (i.e. $D = 2, 4, 6$) of spacetime.

The purpose of our present investigation is to discuss various kinds of discrete and continuous symmetry transformations 
for an {\it odd} dimensional (i.e. $D = 3$) field-theoretic model of a combined system of the  free Abelian 1-form and 
2-form gauge theories and establish that {\it these} symmetry transformations, in their operator form, satisfy a beautiful 
algebra that provides the physical realizations of the abstract mathematical properties (e.g. the algebraic structures, the
operations on a differential form, etc.) that are associated with the de Rham cohomological operators of differential
geometry. To be more precise, our present field-theoretic model is defined on a three $(2 + 1)$-dimensional ($3D$) 
flat Minkowskian spacetime manifold which remains in the background and it does {\it not} participate in  our discussions
(that are connected with the discrete and continuous symmetry transformations). In other words, we focus our discussions 
{\it only} on the {\it internal} symmetries of our BRST-quantized $3D$ theory where any kinds of the discrete 
and/or continuous {\it spacetime} symmetries of the background spacetime manifold are {\it not} considered at all.

We have concentrated, in our present endeavor, on the symmetry properties of the coupled (but equivalent) 
Lagrangian densities [cf. Eqs.~\eqref{1},~\eqref{10} below] that respect {\it total} six infinitesimal and  continuous  
symmetry transformations that include (i)  a set of {\it four} off-shell nilpotent (anti-)BRST and (anti-)co-BRST 
symmetries, (ii) a  {\it unique} bosonic symmetry, and (iii) a ghost-scale symmetry. On top of these continuous symmetries, 
our $3D$ theory {\it also} respects a couple of very useful discrete duality symmetry transformations that are hidden 
in one equation [cf. Eq.~\eqref{6} below]. It has been shown that these symmetry transformation operators obey an algebra 
that is reminiscent of the algebra satisfied by the de Rham cohomological operators of differential geometry 
(cf Section~\ref{sec7} below). We have devoted time on a detailed discussion on the theoretical derivations of the 
Curci-Ferrari (CF) type restrictions on our theory [cf. Eq.~\eqref{22} below] which are the hallmarks of a BRST-quantized theory
and these  restrictions are {\it also} responsible for (i) the absolute anticommutativity of the nilpotent 
(anti-)BRST and (anti-)co-BRST symmetry transformations, (ii) the proof of the uniqueness of the bosonic symmetry transformations, 
and (iii) the existence of the coupled (but equivalent) Lagrangian densities (cf. Secs.~\ref{sec4},~\ref{sec5} below). 
Finally, we have been able to demonstrate the importance of the  algebraic structures that are obeyed by the ghost-scale 
transformation operator  and the  {\it rest} of the transformation operators of our theory 
[cf. Eqs.~\eqref{49},~\eqref{50}] in the context of the consequences that emerge out when the cohomological operators act on 
a given differential form (cf. Secs.~\ref{sec6},~\ref{sec7} below).

The central motivating factors, that have spurred our interest in our present investigation, are as follows. First of all, 
in our previous works (that are connected with the $3D$ {\it combined} system of the free Abelian 1-form and 2-form gauge 
theories as an example of Hodge theory~\cite{rpm6,rpm7})  the discrete duality symmetry transformations 
[cf. Eq.~\eqref{6} below] have {\it not} been discussed at all. Hence, we have {\it not} been able to provide~\cite{rpm6,rpm7}
the physical realization of the Hodge duality operator. We address this issue in our present endeavor. Second, 
in our very recent work~\cite{rpm8},  we have focused {\it only} on the (co-)BRST invariant Lagrangian density 
[cf. Eq.~\eqref{1} below] and have {\it not} considered the anti-BRST and anti-co-BRST invariant Lagrangian density 
[cf. Eq.~\eqref{10} below]. In our present investigation, we concentrate on {\it both} the coupled and equivalent 
Lagrangian densities [cf. Eqs.~\eqref{1},~\eqref{10} below] and discuss their discrete as well as the continuous 
symmetry properties thoroughly. Third, if the beauty of a theory is determined in terms of the numbers
and varieties of symmetry transformations it respects, our present BRST-quantized $3D$ field-theoretic model belongs to this 
class because we show the existence of  a set of {\it six} continuous symmetries as well as a couple of very useful discrete 
duality symmetry transformations for our present $3D$ theory. Fourth, in our earlier works~\cite{rpm6,rpm7}, we have shown 
the validity of the anticommutativity property between an {\it appropriate} pair of nilpotent symmetry transformations 
 (i) by invoking a single CF-type restriction, and (ii) only up to the $U(1)$ gauge symmetry-type 
transformations. In our present endeavor, we establish the absolute anticommutativity property between the 
specific pairs of the nilpotent symmetry transformations by invoking {\it only} the CF-type restrictions. Finally, 
we demonstrate the existence of a two-to-one mapping [cf. Eq.~\eqref{52} below] between the symmetry transformation 
operators and the de Rham cohomological operators which has {\it not} 
been shown in our earlier works~\cite{rpm6,rpm7,rpm8}.

The theoretical contents of our present endeavor are organized as follows. In Section~\ref{sec2}, we
recapitulate the bare essentials of the nilpotent (co-)BRST  transformations that
are respected by an appropriate Lagrangian density~\cite{rpm8}.  Our Section~\ref{sec3} is devoted to the
discussion on the anti-BRST and anti-co-BRST  transformations for the coupled 
(but equivalent) Lagrangian density corresponding to the Lagrangian density of the
previous section. The theoretical contents of our Section~\ref{sec4} are related to the derivations of
the CF-type restrictions on our theory from different theoretical angles. Our Section~\ref{sec5} deals
with the derivation of a {\it unique} set of bosonic transformations which is nothing but the {\it appropriate}
anticommutator of the (anti-)BRST and (anti-) co-BRST transformations. In Section~\ref{sec6}, we devote time on 
the discussion of a set of the ghost-scale symmetry transformation operators and {\it its} algebraic structures with the 
{\it rest} of the {\it continuous} symmetry transformation operators. Our Section~\ref{sec7} contains the {\it complete} algebraic
structures that are obeyed by the symmetry transformation operators of our $3D$ theory and we establish their deep connection 
with the Hodge algebra that is respected by the de Rham cohomological operators of differential geometry\footnote{On a compact spacetime manifold without a boundary, we define {\it three} operators ($d, \, \delta, \,\Delta $) 
which are known as the de Rham cohomological operators of differential geometry~\cite{egh,mm,gss,van,nis,joel} 
where the operator $d = \partial_\mu\, dx^\mu$ (with $d^2 = 0$) is known as the exterior derivative,
the operator $\delta = \pm\, *\, d\, *$ (with $\delta^2 = 0 $) is called as the co-exterior (dual-exterior) 
derivative, and the notation $\Delta = (d + \delta)^2 = \{ d, \; \delta \} $ corresponds to the Laplacian operator. 
In the above relationship (i.e. $\delta = \pm\, *\, d\, *$) between the (co-)exterior derivatives, 
the symbol $*$ stands for the Hodge duality operator on the given compact spacetime manifold. These cohomological operators obey an algebra: $d^2 = \delta^2 =0$, $\{d, \, \delta\} = (d + \delta)^2 = \Delta$, $[\Delta, \, d] = 0$, $[\Delta, \, \delta] = 0$ which is popularly known as the Hodge algebra in the context of differential geometry.}. 
Finally, in Section~\ref{sec8}, we summarize our key results and point out a few future theoretical 
directions that can be pursued later.

In our Appendix~\ref{secA}, we provide an alternative proof of the derivations of the nilpotent (anti-)BRST and (anti-)co-BRST 
transformations for the Nakanishi-Lautrup auxiliary fields [cf. Eq.~\eqref{25} below] which have {\it not} been listed in 
the equations~\eqref{2},~\eqref{3},~\eqref{11} and \eqref{12}. Our Appendix~\ref{secB} is devoted to the derivations of the 
intimate connections between the off-shell nilpotent (anti-)BRST and (anti-)co-BRST symmetry operators by the {\it direct} 
applications of the discrete duality symmetry transformation operators on them.

{\it Notations and Conventions:}
For the flat $3D$ Minkowskian  {\it background} spacetime manifold, we choose the metric tensor $\eta_{\mu\nu} = \text{diag}\,
(+1, -1, -1)$ so that the dot product between two non-null vectors $S_\mu$ and $T_\mu$ is 
$S \cdot T = \eta_{\mu\nu} \,S^\mu\, T^\nu \equiv S_0 \,T_0 - S_i\, T_i$ where the Greek indices 
$\mu, \nu, \sigma, ... = 0, 1, 2$ stand for the time and space directions of our $3D$
spacetime manifold and the Latin indices $i, j, k, ... = 1, 2$ correspond to the space directions {\it only}. We have chosen
the $3D$ Levi-Civita tensor $\varepsilon_{\mu\nu\sigma}$ such that $\varepsilon_{012} = +1 = \varepsilon^{012}$ 
and it satisfies the standard relationships:
$\varepsilon^{\mu\nu\sigma}\varepsilon_{\mu\nu\sigma} = 3!,\;
\varepsilon^{\mu\nu\sigma} \varepsilon_{\mu\nu\eta} = 2! \,\delta^\sigma_\eta, \;
\varepsilon^{\mu\nu\sigma} \varepsilon_{\mu\kappa\eta} = 1! \,
\big(\delta^\nu_\kappa\, \delta^\sigma_\eta - \delta^\nu_\eta \,\delta^\sigma_\kappa \big)$
on the $3D$ flat Minkowskian spacetime manifold (which remains in the background and does {\it not} participate 
in our discussions on the symmetry transformation operators of our theory). We {\it also} adopt the convention of the (i) left derivative w.r.t. {\it all} 
the fermionic fields of our theory, and (ii) derivative w.r.t. the antisymmetric tensor field as
$\big (\partial B_{\mu\nu}/ \partial B_{\sigma \kappa}\big ) 
= \frac{1}{2!} \, (\delta^\sigma_\mu\, \delta^\kappa_\nu - \delta^\sigma_\nu\, \delta^\kappa_\mu)$, 
etc., at the appropriate places in our text.\\


\section{Preliminaries: Nilpotent (co-)BRST Symmetries}
\label{sec2}


Let us begin with the BRST and co-BRST invariant Lagrangian density [${\cal L}_{(B, {\cal B})}$] for the
{\it combined} field-theoretic system of the $3D$ free Abelian 1-form and 2-form gauge theories as (see, e.g.~\cite{rpm8})
\begin{eqnarray}\label{1}
{\cal L}_{(B,{\cal B})} &=& \dfrac{1}{2}\, {\cal B}^\mu {\cal B}_\mu 
- {\cal B}^\mu \Big(\varepsilon_{\mu\nu\sigma} \,\partial^\nu A^\sigma 
- \dfrac{1}{2}\, \partial_\mu \widetilde \phi \Big) - B \,(\partial \cdot A) + \dfrac{B^2}{2}  
+ {\cal B}\, \Big(\dfrac{1}{2}\,\varepsilon_{\mu\nu\sigma} \,\partial^\mu B^{\nu\sigma} \Big) \nonumber\\
&-& \dfrac{{\cal B}^2}{2} + B^\mu \Big( \partial^\nu B_{\nu\mu} - \dfrac{1}{2}\, \partial_\mu \phi \Big) 
- \dfrac{1}{2}\, B^\mu B_\mu - \big(\partial^\mu \bar C^\nu - \partial^\nu \bar C^\mu \big) \big(\partial_\mu C_\nu \big) \nonumber\\
&-& \dfrac{1}{2}\, \Big(\partial  \cdot C - \dfrac{\lambda}{4} \Big) \,\rho 
- \dfrac{1}{2} \,\Big(\partial  \cdot \bar C + \dfrac{\rho}{4} \Big) \,\lambda 
- \dfrac{1}{2}\, \partial^\mu \bar \beta  \,\partial_\mu \beta
- \partial^\mu \bar C  \,\partial_\mu C,
\end{eqnarray}
where the subscripts on the above Lagrangian density denote the abbreviated forms of the
Nakanishi-Lautrup type auxiliary fields $B_\mu$ and ${\cal B}_\mu$ which have been invoked to linearize
the gauge-fixing term for the Abelian 2-form gauge field and kinetic term of the Abelian 1-form
gauge field, respectively. To be precise, total four numbers of Nakanishi-Lautrup type bosonic auxiliary
fields $({\cal B}, \,B,\, {\cal B}_\mu, \,B_\mu)$ have been invoked to linearize the kinetic term
$\big[\text{i.e.}\; \frac{1}{2}\,H^{\mu\nu\sigma}\,H_{\mu\nu\sigma} = \frac{1}{2}\, \big(H_{012} \big)^2 
\equiv \frac{1}{2} \big(\frac{1}{2}\,\varepsilon^{\mu\nu\sigma}\,\partial_\mu B_{\nu\sigma} \big)^2 \big]$
for the Abelian 2-form field, the gauge-fixing term [i.e. $-\,\frac{1}{2}\, (\partial \cdot A)^2$] 
for the Abelian 1-form field, the kinetic term [i.e. $- \frac{1}{2}\, (F_{\mu\nu})^2 \equiv - \frac{1}{2}\, 
(\varepsilon^{\mu\nu\sigma} \, \partial_\nu A_\sigma)^2$]  and the gauge-fixing term 
[i.e. $\frac{1}{2}\, (\partial^\nu B_{\nu\mu}- \frac{1}{2}\, \partial_\mu\phi)^2$] 
for the Abelian 2-form field, respectively. We have the bosonic (pseudo-)scalar fields $(\widetilde \phi)\phi$ 
in our theory due to the reducibility properties of the $3D$ Abelian 1-form and 2-form gauge fields. 
In addition, we have the bosonic set (i.e. $\bar \beta^2 \ne 0,\; \beta^2 \ne 0$) of (anti-)ghost fields 
$(\bar \beta)\beta$ in our theory which are endowed with the ghost numbers $(-2)+2$, respectively. In the 
(co-)BRST invariant Lagrangian density~\eqref{1}, we have the Lorentz vector fermionic 
(i.e. $C^2_\mu  = \bar C^2_\mu = 0, \, C_\mu \, C_\nu + C_\nu \,C_\mu = 0,\;
 C_\mu \, \bar C_\nu + \bar C_\nu \, C_\mu = 0$, etc.) (anti-)ghost fields $(\bar C_\mu) C_\mu$ 
as well as the Lorentz scalar fermionic (i.e. $C^2 = 0,\, \bar C^2 = 0,\, C\, \bar C + \bar C\, C = 0$) 
(anti-)ghost fields $(\bar C)C$ with the ghost numbers $(-1)+1$, respectively. The
{\it fermionic} (i.e. $\rho\, \lambda + \lambda\, \rho = 0 $)
auxiliary (anti-)ghost fields $(\rho)\lambda$ of our system {\it also} carry the ghost numbers 
$(-1)+1$, respectively, because we note that $\rho = -\,2\,(\partial \cdot \bar C)$ and $\lambda = 2\, (\partial \cdot C)$. 
All the above (anti-)ghost fields are required in our theory to maintain the sacrosanct property of unitarity 
which is valid at any arbitrary order of perturbative computation
for {\it all} the physical processes that are allowed by our properly BRST-quantized $3D$ theory.

In our discussions, the Abelian 2-form: $B^{(2)} = \frac{1}{2!}\,B_{\mu\nu}(dx^\mu \wedge dx^\nu)$ 
defines the {\it basic} antisymmetric $(B_{\mu\nu} = - B_{\nu\mu})$ tensor gauge field $B_{\mu\nu}$ and the 3-form: 
$H^{(3)} = d\, B^{(2)} = \frac{1}{3!} \,H_{\mu\nu\sigma} (dx^\mu \wedge dx^\nu \wedge dx^\sigma)$ defines the field-strength
tensor $H_{\mu\nu\sigma} = \partial_\mu B_{\nu\sigma} + \partial_\nu B_{\sigma\mu}+ \partial_\sigma B_{\mu\nu}$
for the antisymmetric basic tensor Abelian gauge field. Here $d = \partial_\mu\,dx^\mu$ 
[with $d^2 = \frac{1}{2!}\,(\partial_\mu \,\partial_\nu - \partial_\nu\, \partial_\mu)(dx^\mu \wedge dx^\nu) = 0$] 
is the exterior derivative of differential geometry~\cite{egh,mm,gss,van,nis,joel}.  Similarly,
for the Abelian 1-form theory, we have the field-strength tensor $F_{\mu\nu} = \partial_\mu A_\nu - \partial_\nu A_\mu$
which is derived from the Abelian 2-form $F^{(2)} = d \,A^{(1)} = \frac{1}{2!}\,F_{\mu\nu}(dx^\mu \wedge dx^\nu)$
where the Abelian 1-form $A^{(1)} = A_\mu \,dx^\mu$ defines the {\it basic} vector gauge field $A_\mu$ of our Abelian gauge theory.

The following infinitesimal, continuous and off-shell nilpotent (i.e. $s^2_d = 0,\, s^2_b = 0$) 
(co-) BRST [i.e. (dual-)BRST] transformations $(s_d)s_b$, namely;
\begin{eqnarray}\label{2}
&& s_d B_{\mu \nu} = \varepsilon_{\mu\nu\sigma}\, \partial^\sigma \bar C, 
\quad s_d A_\mu = - \,\varepsilon_{\mu\nu\sigma}\,\partial^\nu \bar C^\sigma, \quad s_d \bar C_\mu  = - \,\partial_\mu \bar \beta,  \quad s_d  C_\mu  = -\, {\cal B}_\mu,  \nonumber\\
&& s_d  C =  {\cal B}, \quad s_d \beta = -\, \lambda,  \quad s_d \widetilde \phi  = \rho, 
\quad s_d \big[B_\mu, \, {\cal B}_\mu, \, B,\, {\cal B}, \phi, \bar \beta,\, \bar C,\, \rho,  \lambda  \big] = 0,
\end{eqnarray} 
\begin{eqnarray}\label{3}
&& s_b B_{\mu \nu} = - \,\big(\partial_\mu C_\nu - \partial_\nu C_\mu\big),  \quad s_b A_\mu = \partial_\mu C, 
\quad s_b C_\mu  = - \,\partial_\mu \beta, \quad s_b \bar C_\mu  = B_\mu, \nonumber\\
&& s_b \bar C = B, \quad s_b \bar \beta = - \,\rho, \quad s_b \phi  = \lambda, 
\quad s_b \big[B_\mu,\, {\cal B}_\mu, \, B,\, {\cal B}, \widetilde \phi, \beta,\, C,\, \rho, \, \lambda  \big] = 0, 
\end{eqnarray} 
are  the {\it symmetry} transformations
of the action integral: $S = \int d^3 x\, {\cal L}_{(B, {\cal B})}$ because the Lagrangian 
density ${\cal L}_{(B, {\cal B})}$ [cf. Eq.~\eqref{1}] transforms to the total spacetime 
derivatives under the above off-shell nilpotent (co-)BRST symmetry transformations as~\cite{rpm8}:
\begin{eqnarray}\label{4}
s_d {\cal L}_{(B, {\cal B})} &=& - \,\partial_\mu\Big[ \big(\partial^\mu \bar C^\nu 
- \partial^\nu \bar C^\mu \big)\, {\cal B}_\nu 
- {\cal B} \, \partial^\mu \bar C - \frac{1}{2}\, \rho\, {\cal B}^\mu 
- \frac{1}{2}\, \lambda\, \partial^\mu \bar \beta \Big],\\
\nonumber\\
\label{5}
s_b \, {\cal L}_{(B, {\cal B})} &=& - \,\partial_\mu \Big[ \big(\partial^\mu C^\nu 
- \partial^\nu C^\mu \big)\, B_\nu + B \, \partial ^\mu C + \frac{1}{2}\,\lambda\, B^\mu 
- \dfrac{1}{2}\,\rho\, \partial^\mu \beta  \Big].
\end{eqnarray} 
It is interesting to point out that (i) the off-shell nilpotent dual-BRST 
(i.e. co-BRST) symmetry transformations~\eqref{2} are {\it distinctly} different from the corresponding 
transformations in our very recent works~\cite{rpm6,rpm7}   where we have taken $s_d A_\mu = 0$, and 
{ii} the gauge-fixing terms, for the basic gauge fields $A_\mu$ and 
$B_{\mu\nu}$, owe their origin to the co-exterior derivative $\delta = \pm\, *\, d \; *$ of the 
differential geometry~\cite{egh,mm,gss,van,nis,joel}. For instance, it can be readily checked that: 
$\delta \,A^{(1)} = \pm \,*\, d\, * \big(A_\mu\, dx^\mu \big) \equiv \pm\, \big(\partial \cdot A \big)$ 
and $\delta \,B^{(2)} = \pm\, * \,d\, * \big[\frac{1}{2!}\,B_{\mu\nu}\, \big(dx^\mu \wedge dx^\nu\big) \big] 
\equiv \mp \big(\partial^\nu B_{\nu\mu} \big)\, dx^\mu$  where $*$ is the Hodge duality operator defined on 
our chosen $3D$ flat Minkowskian spacetime manifold. It is worthwhile to point out that, in the gauge-fixing 
term for the Abelian 2-form field $B_{\mu\nu}$, the pure  scalar field $\phi$ (with proper mass dimension)  
appears due to the stage-one reducibility in the theory.

In addition to the above off-shell nilpotent (co-)BRST symmetry transformations, the Lagrangian 
density~\eqref{1} respects a set of following discrete duality symmetry transformations:
\begin{eqnarray}\label{6}
&& A_\mu \longrightarrow \,\mp\, \dfrac{i}{2}\, \varepsilon_{\mu\nu\sigma}\, B^{\nu\sigma},  \qquad\quad  
B_{\mu\nu} \longrightarrow \,\pm i\,\varepsilon_{\mu\nu\sigma}\, A^\sigma, \nonumber\\
&& B \longrightarrow \, \mp\, i {\cal B}, \quad \qquad {\cal B} \longrightarrow \,\pm\, i B,
\qquad B_\mu \longrightarrow \,\mp\, i \,{\cal B}_\mu, \quad \quad {\cal B}_\mu \longrightarrow\, \; \pm\, i\, B_\mu,  \nonumber\\
&&  \phi \longrightarrow \, \mp\, i\, \widetilde \phi, \qquad \quad\; \widetilde \phi \longrightarrow \,\pm\, i \phi, 
\qquad C_\mu \,\longrightarrow\, \pm\,i\,\bar C_\mu, \quad \quad  \bar C_\mu \,\longrightarrow\, \pm\,i\, C_\mu, \nonumber\\
&& C \,\longrightarrow\, \mp\,i\,\bar C, \qquad \quad \bar C \,\longrightarrow\, \mp\,i\, C,  
\qquad \beta \,\longrightarrow\, \pm\,i\,\bar \beta, \;\;\qquad \quad \bar \beta \,\longrightarrow\, \mp\,i\,\beta, \quad \qquad \nonumber\\
&& \rho \,\longrightarrow\, \mp\,i\,\lambda, \;\;\qquad\quad \lambda \,\longrightarrow\, \mp\,i\,\rho. 
\end{eqnarray}
It is interesting to point out that, in the above symmetry transformations, the {\it basic} Abelian 1-form and 2-form
fields are connected with each-other. In fact, these fields are dual to each-other in the language of the Hodge duality
$*$ operator that is defined on the flat $3D$ Minkowskian spacetime manifold (see, e.g.~\cite{rpm8} for details).

Before we conclude this section, we would like to remark that the off-shell nilpotent (co-)BRST symmetry 
transformation operators [cf. Eqs.~\eqref{2},~\eqref{3}] and the discrete duality symmetry transformation operators 
[cf. Eq.~\eqref{6}] are interconnected with one-another by a beautiful  
mathematical relationship. To be precise, we note that the following equation
\begin{eqnarray}\label{7}
&& s_d \, \Phi = \pm\, *\, s_b \, *\, \Phi, \nonumber\\
&& \Phi = B_{\mu\nu},\, A_\mu, \, B_\mu, \,{\cal B}_\mu, \, \bar C_\mu, \, C_\mu, \, \phi, \, \widetilde \phi,\,
B,\,  {\cal B},\,  \bar \beta,\,  \beta, \, \bar C, \, C,\, \rho, \,\lambda,  
\end{eqnarray}
provides an intimate connection amongst the continuous and nilpotent symmetry transformation operators $s_d$ 
and $s_b$ and the discrete duality symmetry transformations because the symbol $*$ stands for the 
transformation operators~\eqref{6} and, as is obvious, the generic field $\Phi$ denotes {\it all} the fields 
that are present in the Lagrangian density~\eqref{1}. We would like to point out that the ($\pm$) signs, 
on the r.h.s. of the above equation, are decided and dictated by a couple of successive operations of the 
discrete duality symmetry transformation operators~\eqref{6} on the generic field $\Phi$ which incorporates the 
bosonic fields ($\mathbb{B}$) as well as the fermionic fields ($\mathbb{F}$). In other words, the 
following operator equations, namely;
\begin{eqnarray}\label{8}
 * \, \big (*\, \Phi \big ) = \pm\, \Phi, \qquad  * \, \big (*\,\mathbb{B} \big ) = +\, \mathbb{B}, 
\qquad  * \, \big (*\, \mathbb{F} \big ) = -\, \mathbb{F},
\end{eqnarray}
demonstrate that we have the {\it plus} sign for the bosonic fields ($\mathbb{B}$) and {\it negative} sign for the 
fermionic fields ($\mathbb{F}$) in~\eqref{7}. These fields are present in the  (co-)BRST invariant Lagrangian 
density~\eqref{1} of our $3D$ field-theoretic model which incorporates the combined system of the BRST-quantized
free Abelian 1-form and 2-form gauge theories. It is extremely interesting to point out that
a {\it reciprocal} operator relationship {\it also} exists in our theory, namely;
\begin{eqnarray}\label{9}
&& s_b \, \Phi = \mp\, *\, s_d \, *\, \Phi, \nonumber\\ 
&& \Phi = B_{\mu\nu},\, A_\mu,\,  B_\mu,\, {\cal B}_\mu, \, \bar C_\mu,\,  C_\mu, \, \phi, \, \widetilde \phi,\, 
B, \, {\cal B}, \, \bar \beta, \, \beta,\, \bar C, \, C,\, \rho,\, \lambda,  
\end{eqnarray}
which is the analogue of mathematical operator relationship in~\eqref{7} where the signs ($\mp $), 
on the r.h.s., are decided and dictated by the relationship in~\eqref{8}, once again. It is clear, 
however, that the {\it negative} sign is associated with the bosonic fields ($\mathbb{B}$) and {\it plus} 
sign is related to  the fermionic fields ($\mathbb{F}$) of our theory in~\eqref{9}.

We end this section with the following remarks. First of all, we note that the kinetic terms for the 
Abelian 1-form and 2-form gauge fields, owing their origin to the exterior derivative $d$ of 
differential geometry, remain invariant under the BRST symmetry transformations. On the other hand, 
it is the gauge-fixing terms of the {\it above} fields, owing their origin to 
the co-exterior derivative $\delta =  \pm\, *\, d\, *$ of differential geometry, that are found to remain 
{\it unchanged} under the co-BRST symmetry transformations. Second, it is very interesting to point out that, 
under the discrete duality symmetry transformations~\eqref{6}, the ghost-sector and non-ghost sectors of 
the (co-)BRST invariant Lagrangian density (1) remain invariant separately and independently. Finally, we 
observe that the off-shell nilpotent (i.e. $s_d^2 = 0, \, s_b^2 = 0 $) versions of the (co-)BRST symmetry
transformations [cf. Eqs.~\eqref{2},~\eqref{3}] do {\it not} anticommute with each-other 
(i.e. $\{ s_d, \; s_b \} \neq 0$). In fact, their anticommutator leads to the definition of a bosonic 
symmetry transformation operator in our $3D$  theory (cf. Section~\ref{sec5} below).


\section{Nilpotent Anti-BRST and Anti-co-BRST Symmetries: Coupled Lagrangian Densities }
\label{sec3}


Analogous to the (co-)BRST invariant Lagrangian density~\eqref{1}, we can write down the anti-BRST 
and anti-co-BRST invariant {\it coupled} Lagrangian density as follows 
\begin{eqnarray}\label{10}
{\cal L}_{(\bar B, \bar {\cal B})} &=& \dfrac{1}{2}\, \bar {\cal B}^\mu \bar {\cal B}_\mu 
+ \bar {\cal B}^\mu \Big(\varepsilon_{\mu\nu\sigma} \,\partial^\nu A^\sigma 
+ \dfrac{1}{2}\, \partial_\mu \widetilde \phi \Big) - B \,(\partial \cdot A)+ \dfrac{B^2}{2}  
+ {\cal B}\, \Big(\dfrac{1}{2}\,\varepsilon_{\mu\nu\sigma} \,\partial^\mu B^{\nu\sigma} \Big) \nonumber\\
&-& \dfrac{{\cal B}^2}{2} - \bar B^\mu \Big( \partial^\nu B_{\nu\mu} + \dfrac{1}{2}\, \partial_\mu \phi \Big)
- \dfrac{1}{2}\, \bar B^\mu \bar B_\mu 
- \big(\partial^\mu \bar C^\nu - \partial^\nu \bar C^\mu \big) \big(\partial_\mu C_\nu \big) \nonumber\\
&-& \dfrac{1}{2}\, \Big(\partial  \cdot C - \dfrac{\lambda}{4} \Big)\, \rho 
- \dfrac{1}{2} \,\Big(\partial  \cdot \bar C + \dfrac{\rho}{4} \Big) \,\lambda 
- \dfrac{1}{2}\, \partial^\mu \bar \beta  \,\partial_\mu \beta - \partial^\mu \bar C  \,\partial_\mu C,
\end{eqnarray}
where the subscripts $(\bar B, \bar {\cal B}) $ on the above Lagrangian density denote the abbreviated forms 
of the Nakanishi-Lautrup type auxiliary fields $\bar B_\mu$ and $\bar {\cal B}_\mu $ which have been invoked 
to linearize the gauge-fixing term for the Abelian 2-form field and kinetic term for the Abelian 1-form field, 
respectively. A few noteworthy points, at this stage, are as follows. First of all, we note that the ghost-sector 
of the {\it above} Lagrangian density is  exactly {\it same} as in the Lagrangian density~\eqref{1}. Second, the bosonic 
auxiliary fields (i.e. $B, \, {\cal B} $), invoked to linearize the gauge-fixing term for the 
Abelian 1-form field and kinetic term for the $3D$ Abelian 2-form field, respectively, remain the {\it same} in~\eqref{1} and 
\eqref{10}. Third, the bosonic Lorentz vector auxiliary fields (i.e. $\bar B_\mu, \; \bar {\cal B}_\mu $) 
in~\eqref{10} are {\it different} from the auxiliary fields (i.e. $B_\mu, \; {\cal B}_\mu $) in~\eqref{1} because the 
signs of the (pseudo-)scalar fields $(\widetilde \phi)\phi$ have been changed in~\eqref{10} for the sake of 
generality of the kinetic term for the Abelian 1-form field and gauge-fixing term for the Abelian 2-form field, 
respectively. Finally, the Lagrangian densities~\eqref{1} and \eqref{10} are {\it coupled} because the pairs of 
Nakanishi-Lautrup auxiliary fields ($B_\mu, \, {\cal B}_\mu$) and ($\bar B_\mu, \; \bar {\cal B}_\mu$) are {\it connected} 
with each-other by the CF-type restrictions (cf. Section~\ref{sec4} below).

It is straightforward to check that the following off-shell nilpotent (i.e. $s_{ab}^2 = 0, \, s_{ad}^2 = 0$), 
infinitesimal and continuous anti-BRST ($s_{ab} $) and anti-co-BRST ($s_{ad}$) transformations 
\begin{eqnarray}\label{11}
&& s_{ab} B_{\mu \nu} = -\big(\partial_\mu \bar C_\nu - \partial_\nu \bar C_\mu\big), 
\quad s_{ab} A_\mu = \partial_\mu \bar C, \quad s_{ab} \bar C_\mu  = - \partial_\mu \bar \beta, 
\quad s_{ab}  C_\mu  = \bar B_\mu, \nonumber\\
&&  s_{ab}  C = - B, \quad  s_{ab}  \beta = - \lambda, \quad s_{ab} \phi  = \rho, 
\quad  s_{ab} \big[{\bar B}_\mu, \, \bar {\cal B}_\mu, \, B, \, {\cal B}, \, \widetilde \phi, \, 
\bar\beta,\, \bar C,\, \rho, \, \lambda \big] = 0, \qquad \\
\nonumber\\
\label{12}
&& s_{ad} B_{\mu \nu} = + \varepsilon_{\mu\nu\sigma}\, \partial^\sigma  C, \quad 
s_{ad} A_\mu = - \varepsilon_{\mu\nu\sigma}\,\partial^\nu  C^\sigma, 
\quad s_{ad}  C_\mu  =  \partial_\mu \beta,   \quad s_{ad} \bar  C_\mu  = - \bar {\cal B}_\mu, \nonumber\\
&& s_{ad} \bar C = - {\cal B}, \quad s_{ad} \bar \beta = \rho, \quad s_{ad} \widetilde \phi  = \lambda, 
\quad  s_{ad} \big[{\bar B}_\mu, \, \bar {\cal B}_\mu, \,B,\, {\cal B}, \,\phi,\, \beta,\, C,\, \rho, \, \lambda\big] = 0, \qquad
\end{eqnarray}
are the {\it symmetry} transformations of the action integral 
($S = \int d^3 x\,{\cal L}_{(\bar B, \bar {\cal B})} $) because we observe that the 
Lagrangian density~\eqref{10} transforms to the  total spacetime derivatives as 
\begin{eqnarray}\label{13}
s_{ab} {\cal L}_{(\bar B, \bar {\cal B})} &=& \partial_\mu\Big[\big(\partial^\mu  \bar C^\nu 
- \partial^\nu  \bar C^\mu \big) \bar B_\nu - B \,\partial^\mu  \bar C  
- \dfrac{1}{2}\, \rho\, {\bar B}^\mu + \dfrac{1}{2}\, \lambda\, \,\partial^\mu  \bar \beta  \Big], \\
&&\nonumber\\ 
\label{14}
s_{ad} {\cal L}_{(\bar B, \bar {\cal B})} &=& \partial_\mu\Big[\big(\partial^\mu  C^\nu 
- \partial^\nu  C^\mu \big) \bar {\cal B}_\nu + {\cal B} \, \partial^\mu  C 
+ \dfrac{1}{2}\, \lambda\, \bar {\cal B}^\mu - \dfrac{1}{2}\, \rho \, \partial^\mu  \beta  \Big],
\end{eqnarray}
thereby rendering the {\it above} action integral invariant due to Gauss's divergence theorem (because of 
which all the fields vanish off as $x\to \pm\,\infty $). It is very interesting to point out that the 
coupled Lagrangian density~\eqref{10} {\it also} respects a set of discrete duality symmetry transformations 
which is the {\it analogue} of the set of duality symmetry transformations~\eqref{6} such that all the discrete symmetry 
transformations for all the fields are {\it same} as in \eqref{6} except the replacements: 
$ \bar B_\mu \longrightarrow \mp\, i \bar {\cal B}_\mu, \; \bar {\cal B}_\mu \longrightarrow \pm\, i \bar B_\mu$
in the place of $ B_\mu \longrightarrow \mp\, i  {\cal B}_\mu, \; {\cal B}_\mu \longrightarrow \pm\, i  B_\mu $. 
The interplay between the off-shell nilpotent continuous symmetry transformations in~\eqref{11} and \eqref{12} 
and discrete duality symmetry transformations  for the the Lagrangian density~\eqref{10} provide the analogues of 
the operator relationships in~\eqref{7} and \eqref{9} as follows
\begin{eqnarray}\label{15}
 s_{ad} \,  \widetilde {\Phi} = \pm\, *\, s_{ab} \, *\, \widetilde {\Phi}, \qquad 
s_{ab} \, \widetilde {\Phi} = \mp\, *\, s_{ad} \, *\, \widetilde {\Phi},
\end{eqnarray}
where the signs, on the r.h.s. of {\it both} the entries in the above equation, are dictated by a couple 
of successive operations of the discrete duality symmetry transformations on the generic field $\widetilde {\Phi} $ 
of  the Lagrangian density \eqref{10}. In other words, we have the explicit form 
of $\widetilde {\Phi} = B_{\mu\nu}, A_\mu,  \bar B_\mu, \bar {\cal B}_\mu,  \bar C_\mu,  C_\mu,  \phi,  \widetilde \phi,
B,  {\cal B},  \bar \beta,  \beta, \bar C,  C, \rho, \lambda$.
It is obvious that the symbol $*$, in the above equation, corresponds 
to the discrete duality symmetry transformations for the anti-BRST and anti-co-BRST invariant 
Lagrangian density ${\cal L}_{(\bar B, \bar {\cal B})}$ [cf. Eq.~\eqref{10}].

We conclude this section with the following crucial remarks. First of all, we note that the kinetic terms
for the Abelian 1-form and 2-form gauge fields, owing their origin to the exterior derivative $d$ of 
differential geometry, remain invariant under the anti-BRST symmetry transformations. On the other hand, 
we observe that the total gauge-fixing terms for the Abelian 1-form and 2-form fields, owing their origin 
{\it primarily} to the co-exterior derivative $\delta = \pm \, *\, d\, * $, remain unchanged under the nilpotent 
anti-co-BRST symmetry transformations. Second, under the discrete duality symmetry transformations
for  the Lagrangian density ${\cal L}_{(\bar B, \bar {\cal B})}$ [cf. Eq.~\eqref{10}], the (non-)ghost sectors 
of our theory remain invariant separately and independently. Third, we note that the (co-)BRST invariant 
Lagrangian density~\eqref{1} can be expressed as the sum of (i) the kinetic terms for the Abelian 1-form 
and 2-form gauge fields, and (ii) the terms that can be written in the language of the
(anti-)BRST symmetry transformations ($s_{(a)b} $) [cf. Eqs.~\eqref{11},~\eqref{3}] as follows:
\begin{eqnarray}\label{16}
{\cal L}_{(B, {\cal B})} &=& \dfrac{1}{2}\, {\cal B}^\mu {\cal B}_\mu 
- {\cal B}^\mu \Big(\varepsilon_{\mu\nu\sigma} \, \partial^\nu A^\sigma 
- \dfrac{1}{2}\, \partial_\mu \widetilde \phi \Big) 
+ {\cal B} \Big(\dfrac{1}{2}\,\varepsilon_{\mu\nu\sigma}\, \partial^\mu B^{\nu\sigma} \Big) - \dfrac{{\cal B}^2} {2} \nonumber\\
&+& s_b\,s_{ab} \Big[\dfrac{1}{4}\, B^{\mu\nu} \,B_{\mu\nu} + \dfrac{1}{2}\, A^\mu\, A_\mu 
+ \dfrac{1}{4} \,\phi^2 - \dfrac{1}{4}\, \widetilde \phi^2  + \dfrac{1}{4}\, \bar \beta\, \beta 
- \dfrac{1}{2}\, \bar C^\mu\, C_\mu + \dfrac{1}{2}\, \bar C\,C   \Big]. \qquad
\end{eqnarray} 
The above form of the Lagrangian density is important in proving the BRST invariance of 
${\cal L}_{(B, {\cal B})}$ because of (i) the off-shell nilpotency (i.e. $s_b^2 = 0$) property of 
the BRST symmetry transformations~\eqref{3}, and (ii) the BRST-invariance of the kinetic terms of 
the Abelian 1-form and 2-form gauge fields because of our observations: 
$s_b {\cal B}_\mu = 0, \; s_b {\cal B} = 0, \; s_b \widetilde \phi = 0,\;
s_b A_\mu = \partial_\mu C, \; s_b B_{\mu\nu} = - (\partial_\mu C_\nu - \partial_\nu C_\mu) $ 
in the equation \eqref{3}. Fourth, the Lagrangian density~\eqref{1} can be {\it also} 
expressed in terms of the off-shell nilpotent  (anti-)co-BRST symmetry transformations~\eqref{12} 
and \eqref{2} as follows
\begin{eqnarray}\label{17}
{\cal L}_{(B, {\cal B})} &=& B^\mu \Big(\partial^\nu B_{\nu\mu} - \dfrac{1}{2}\partial_\mu \phi \Big) 
-  \dfrac{1}{2}\, B^\mu \,B_\mu - B \, \big(\partial \cdot A \big) + \dfrac{B^2}{2} \nonumber\\
&+& s_d\,s_{ad} \Big[-\dfrac{1}{4}\, B^{\mu\nu} \,B_{\mu\nu} - \dfrac{1}{2}\, A^\mu\, A_\mu + \dfrac{1}{4} \,\phi^2 
- \dfrac{1}{4}\, \widetilde \phi^2 + \dfrac{1}{4}\, \bar \beta\, \beta - \dfrac{1}{2}\, \bar C^\mu\, C_\mu 
+ \dfrac{1}{2}\, \bar C\,C   \Big], \qquad
\end{eqnarray} 
which turns out to be the sum of the gauge-fixing terms for the Abelian 2-form as well as 1-form gauge 
fields and a co-BRST {\it exact} quantity. The above form is important in the sense that the co-BRST 
insurance of the Lagrangian density ${\cal L}_{(B, {\cal B})}$ [cf. Eq.~\eqref{1}] can be proven
in a straightforward manner due to our observations that (i) the co-BRST symmetry transformations~\eqref{2} 
are off-shell nilpotent (i.e. $s_d^2 = 0 $), and (ii) the gauge-fixing terms of the Abelian 1-form and 2-form 
gauge fields remain invariant under the co-BRST symmetry transformations~\eqref{2} because
of our observations: $s_d B_{\mu \nu} = \varepsilon_{\mu\nu\sigma}\, \partial^\sigma \bar C,\; s_d A_\mu 
= - \,\varepsilon_{\mu\nu\sigma}\, \partial^\nu \bar C^\sigma, \; s_d \phi = 0, \; s_d B_\mu = 0,\; s_d B = 0$. 
Fifth, the analogues of the equations~\eqref{16} and \eqref{17} can be {\it also}
written for the Lagrangian density~\eqref{10} as
\begin{eqnarray}\label{18}
{\cal L}_{(\bar B, \bar {\cal B})}&=& \dfrac{1}{2}\, \bar {\cal B}^\mu\, \bar {\cal B}_\mu 
+ \bar {\cal B}^\mu \,\Big(\varepsilon_{\mu\nu\sigma} \, \partial^\nu A^\sigma 
+ \dfrac{1}{2}\, \partial_\mu \widetilde \phi \Big) 
+ {\cal B}\, \Big(\dfrac{1}{2}\,\varepsilon_{\mu\nu\sigma}\, \partial^\mu B^{\nu\sigma} \Big) 
- \dfrac{{\cal B}^2} {2} \nonumber\\
&-& s_{ab}\,s_b \Big[\dfrac{1}{4}\, B^{\mu\nu} B_{\mu\nu} + \dfrac{1}{2}\, A^\mu A_\mu 
+ \dfrac{1}{4} \,\phi^2 - \dfrac{1}{4}\, \widetilde \phi^2 + \dfrac{1}{4}\, \bar \beta\, \beta 
- \dfrac{1}{2}\, \bar C^\mu\, C_\mu + \dfrac{1}{2}\, \bar C\,C   \Big], \quad
\end{eqnarray}
\begin{eqnarray}\label{19}
{\cal L}_{(\bar B, \bar {\cal B})} &=& -\bar B^\mu \Big(\partial^\nu B_{\nu\mu} 
+ \dfrac{1}{2}\partial_\mu \phi \Big) -  \dfrac{1}{2}\, \bar B^\mu \,\bar B_\mu 
- B \, \big(\partial \cdot A \big) + \dfrac{B^2}{2} \nonumber\\
&-& s_{ad}\, s_d \Big[-\dfrac{1}{4}\, B^{\mu\nu} B_{\mu\nu} - \dfrac{1}{2}\, A^\mu A_\mu 
+ \dfrac{1}{4} \,\phi^2 - \dfrac{1}{4}\, \widetilde \phi^2 + \dfrac{1}{4}\, \bar \beta\, \beta 
- \dfrac{1}{2}\, \bar C^\mu\, C_\mu + \dfrac{1}{2}\, \bar C\,C   \Big], \qquad 
\end{eqnarray}
which are in such a mathematically nice forms that the anti-BRST and anti-co-BRST invariance can be proven in a 
straightforward manner because of (i) the off-shell nilpotency (i.e. $s_{ab}^2 = 0, \, s_{ad}^2 = 0 $) 
properties of the anti-BRST symmetry transformations ($s_{ab}$) and anti-co-BRST symmetry
transformations ($s_{ad}$) [cf. Eqs.~\eqref{11},~\eqref{12}],  (ii) the invariance of the kinetic 
terms of the Abelian 1-form and 2-form gauge fields under the anti-BRST symmetry transmissions, and 
(iii) the invariance of the gauge-fixing terms for the Abelian 1-form and 2-form gauge fields under the
anti-co-BRST symmetry transformations.  Finally, in the proofs of \eqref{16},~\eqref{17},~\eqref{18}  and~\eqref{19}, 
one has to use (i) the off-shell nilpotent symmetry transformations \eqref{25}, (ii) the validity of 
the CF-type restrictions: $B_\mu + \bar B_\mu + \partial_\mu \phi = 0, \;
{\cal B}_\mu + \bar {\cal B}_\mu + \partial_\mu \widetilde \phi = 0$ (cf. Section~\ref{sec4} below for details), and (iii) the
anticommutativity properties of the basic and auxiliary fermionic fields of our theory. In other words, 
the CF-type restrictions [cf. Eq.~\eqref{22} below] and their (anti-)BRST and (anti-)co-BRST symmetry invariance
[cf. Eq.~\eqref{25} below] along with the anticommutativity  properties of the fermionic fields 
have been taken into account as far as the proofs of 
the precise forms of the Lagrangian densities [cf. Eqs.~\eqref{16}, \eqref{17},~\eqref{18},~\eqref{19}]
are concerned (modulo some total spacetime derivatives).\\


\section{Curci-Ferrari Type Restrictions: Explicit Derivations}
\label{sec4} 
 
Our present section is divided into {\it three} subsections where we derive the (anti-)BRST and 
(anti-)co-BRST invariant  Curci-Ferrari (CF) type restrictions whose existence is one of the hallmarks 
of the BRST-quantized gauge and/or reparameterization  invariant theories~\cite{bon1,bon2}.  
In Subsect.~\ref{sec4.1}, by demanding the {\it direct} equality of the  Lagrangian densities~\eqref{1} and~\eqref{10}, 
we demonstrate the  existence of the CF-type restrictions on our theory. Our Subsect.~\ref{sec4.2} deals with the 
derivations of the CF-type restrictions from the point of view of the symmetry considerations of the 
Lagrangian densities~\eqref{1} and~\eqref{10}. In Subsect.~\ref{sec4.3}, the requirements of the absolute 
anticommutativity between the off-shell nilpotent  and continuous (i)  (anti-)BRST symmetry transformation operators, 
and (ii) anti-)co-BRST symmetry transformation operators, lead to the validity of the CF-type restrictions.


\subsection{Direct Equality of ${\cal L}_{(B, {\cal B})} $ and ${\cal L}_{(\bar B, \bar {\cal B})}$: CF-Type Restrictions}
\label{sec4.1}


Both the Lagrangian densities~\eqref{1} and~\eqref{10} are coupled and equivalent due to the presence of a 
set of CF-type restrictions on our theory. In our present section, we explain this fact in a very lucid manner. To be
precise, we show that ${\cal L}_{(B, {\cal B})} - {\cal L}_{(\bar B, \bar {\cal B})} = 0 $ is {\it true} if and 
only if the CF-type restrictions are satisfied. Thus, both the Lagrangian densities~\eqref{1} and~\eqref{10} 
are {\it equivalent} on the submanifold of fields where the CF-type restrictions are valid. To corroborate 
this statement, let us note that (i) the {\it total} FP-ghost part, (ii) the kinetic term for the $3D$ Abelian 
2-form field, and (ii) the gauge-fixing term for the Abelian 1-form gauge field are {\it common} in both the Lagrangian 
densities. As a consequence, they cancel out in the {\it above} difference. Thus, ultimately, we have the 
following  expression for the explicit difference between the Lagrangian densities~\eqref{1} and~\eqref{10}, namely;
\begin{eqnarray}\label{20}
{\cal L}_{(B, {\cal B})} - {\cal L}_{(\bar B, \bar {\cal B })} &=& \dfrac{1}{2}\, {\cal B}^\mu {\cal B}_\mu 
- {\cal B}^\mu \Big(\varepsilon_{\mu\nu\sigma} \,\partial^\nu A^\sigma - \dfrac{1}{2}\, \partial_\mu \widetilde \phi \Big) 
+ B^\mu \Big( \partial^\nu B_{\nu\mu} - \dfrac{1}{2}\, \partial_\mu \phi \Big) 
-\dfrac{1}{2}\, B^\mu B_\mu\nonumber\\
&-& \dfrac{1}{2}\, \bar {\cal B}^\mu \bar {\cal B}_\mu - \bar {\cal B}^\mu \Big(\varepsilon_{\mu\nu\sigma} \,\partial^\nu A^\sigma 
+ \dfrac{1}{2}\, \partial_\mu \widetilde \phi \Big) + \bar B^\mu \Big( \partial^\nu B_{\nu\mu} 
+ \dfrac{1}{2}\, \partial_\mu \phi \Big) + \dfrac{1}{2}\, \bar B^\mu \bar B_\mu.\nonumber\\
\end{eqnarray}
The above expression can be further  simplified by applying the straightforward 
algebraic tricks to recast it as a sum of the total spacetime derivative terms {\it plus} a couple of 
specific form of terms in their factorized forms as follows:
\begin{eqnarray}\label{21}
{\cal L}_{(B, {\cal B})} - {\cal L}_{(\bar B, \bar {\cal B })}  &=& \partial_\mu\,
\Big[\varepsilon^{\mu\nu\sigma}\,\widetilde \phi\, (\partial_\nu A_\sigma) - \phi \,(\partial_\nu B^{\nu \mu}) \Big] \nonumber\\
&-& \Big(\varepsilon^{\mu\nu\sigma}\, \partial_\nu A_\sigma - \dfrac{1}{2} \,{\cal B}^\mu 
+ \dfrac{1}{2}\, \bar {\cal B}^\mu \Big) \Big[{\cal B}_\mu + \bar {\cal B}_\mu + \partial_\mu \widetilde \phi \Big] \nonumber\\
&+& \Big(\partial_\nu B^{\nu\mu} - \dfrac{1}{2}\,B^\mu 
+ \dfrac{1}{2}\, \bar B^\mu \Big) \Big[B_\mu + \bar B_\mu + \partial_\mu \phi \Big].
\end{eqnarray}
A close and careful look at the above equation demonstrates that {\it both} the Lagrangian densities~\eqref{1} 
and~\eqref{10} are {\it equivalent} modulo a total spacetime derivative (which does not play any significant 
role in the description of the dynamics of our theory) and the following CF-type restrictions on our theory, namely; 
 \begin{eqnarray}\label{22}
B_\mu + \bar B_\mu + \partial_\mu \phi = 0, \qquad \quad
{\cal B}_\mu + \bar {\cal B}_\mu + \partial_\mu \widetilde \phi = 0.
\end{eqnarray}
It is very interesting to point out that the following EL-EoMs 
 \begin{eqnarray}\label{23}
&&{\cal B}_\mu = \varepsilon_{\mu\nu\sigma} \,\partial^\nu A^\sigma 
- \dfrac{1}{2}\, \partial_\mu \widetilde \phi, \qquad\;\;  B_\mu 
= \partial^\nu B_{\nu\mu} - \dfrac{1}{2}\, \partial_\mu \phi, \nonumber\\
&&\bar {\cal B}_\mu = - \varepsilon_{\mu\nu\sigma} \partial^\nu A^\sigma 
- \dfrac{1}{2}\, \partial_\mu \widetilde \phi, \qquad  \bar B_\mu 
= - \partial^\nu B_{\nu\mu} - \dfrac{1}{2}\, \partial_\mu \phi, 
\end{eqnarray} 
which are derived from the Lagrangian densities~\eqref{1} and \eqref{10}, w.r.t. the Nakanishi-Lautrup auxiliary fields 
(${\cal B}_\mu, \, B_\mu, \, \bar {\cal B}_\mu, \, \bar B_\mu$), also imply the above CF-type restrictions~\eqref{22}. 
However, we would like to lay the emphasis on the fact that the CF-type restrictions {\it themselves} are {\it not} the
EL-EoMs because they are {\it not} derived from any {\it specific} single Lagrangian density.

The existence of the CF-type restrictions [cf. Eq.~\eqref{22}]  (i) establishes that the Lagrangian 
densities~\eqref{1} and \eqref{10} are the specific set of {\it coupled} Lagrangian densities, and (ii) is the 
hallmark (see, e.g.~\cite{bon1,bon2} for details) of our BRST-quantized $3D$ combined system 
of the free Abelian 1-form and 2-form gauge theories. The {\it latter} observation implies that these restrictions 
[cf. Eq.~\eqref{22}] are the unavoidable {\it physical} restrictions on our theory and, hence, 
they must be (anti-)BRST as well as (anti-)co-BRST invariant, namely;   
\begin{eqnarray}\label{24}
&& s_{(a)b} \Big[B_\mu + \bar B_\mu + \partial_\mu \phi  \Big] = 0, \qquad
s_{(a)b} \Big[{\cal B}_\mu + \bar {\cal B}_\mu + \partial_\mu \widetilde \phi  \Big] = 0, \nonumber\\
&& s_{(a)d} \Big[B_\mu + \bar B_\mu + \partial_\mu \phi  \Big] = 0, \qquad
s_{(a)d} \Big[{\cal B}_\mu + \bar {\cal B}_\mu + \partial_\mu \widetilde \phi  \Big] = 0.
\end{eqnarray}  
The above sacrosanct requirements lead to the following {\it additional} (anti-)BRST as well as 
(anti-)co-BRST symmetry transformations for the Nakanishi-Lautrup fields
\begin{eqnarray}\label{25}
&& s_d \bar {\cal B}_\mu = - \partial_\mu \rho, \qquad s_d {\bar B}_\mu = 0,
\qquad s_b \bar B_\mu = - \partial_\mu \lambda, \qquad\;\;  s_b \bar {\cal B}_\mu =0,  \nonumber\\
&& s_{ab} B_\mu = - \partial_\mu \rho, \qquad  s_{ab}  {\cal B}_\mu =0, \qquad
s_{ad} {\cal B}_\mu = - \partial_\mu \lambda, \qquad s_{ad} B_\mu = 0,
\end{eqnarray} 
 which have {\it not} been listed in the equations~\eqref{2},~\eqref{3},~\eqref{11} and~\eqref{12}. 
However, these symmetry  transformations (i) are off-shell nilpotent of order two, and (ii) are very 
important because they will play very crucial roles in our next subsection where we  shall derive the 
(anti-)BRST and (anti-)co-BRST invariant CF-type restrictions from the symmetry considerations.

 We end this subsection with the following remarks. First of all, we have observed that our 
$3D$ combined system of the free Abelian 1-form  and 2-form gauge theories is endowed with a 
couple of {\it non-trivial} CF-type restrictions (i.e. $B_\mu + \bar B_\mu + \partial_\mu \phi = 0, \;
{\cal B}_\mu + \bar {\cal B}_\mu + \partial_\mu \widetilde \phi = 0$) which are the hallmarks 
of our present $3D$ BRST-quantized theory (see, e.g.~\cite{bon1,bon2} for details). 
Second, it is the presence of these CF-type restrictions that we have been able to prove that our 
Lagrangian densities~\eqref{1} and \eqref{10} are coupled and equivalent. Third, the CF-type restrictions 
are {\it physical} restrictions on our theory because they are (anti-)BRST as well as (anti-)co-BRST 
invariant [cf. Eq.~\eqref{24}] under the symmetry transformations \eqref{2},~\eqref{3},~\eqref{11},~\eqref{12} 
and~\eqref{25}. Fourth, as stated earlier, the CF-type restrictions~\eqref{22} are {\it not} the EL-EoMs 
because they do {\it not} emerge out  either from Lagrangian density~\eqref{1} or from \eqref{10} separately 
and independently. Finally, it is clear from equation~\eqref{23} that the
restrictions: $ B_\mu - \bar B_\mu = 2 \, (\partial^\nu B_{\nu\mu}), \; {\cal B}_\mu - \bar {\cal B}_\mu 
= 2 \, \varepsilon_{\mu\nu\sigma}\, \partial^\nu A^\sigma $ are also true and they can satisfy~\eqref{21}.
However, these are {\it not} chosen as the CF-type restrictions because of other reasons which
are explained in our forthcoming Subsecs.~\ref{sec4.2} and~\ref{sec4.3} where we take into account the 
symmetry considerations for the coupled Lagrangian densities and the requirements of the absolute anticommutativity
properties between the specific set of the off-shell nilpotent and continuous  symmetry transformation operators
of our $3D$ BRST-quantized field-theoretic system.


\subsection{CF-Type Restrictions: Symmetry Considerations}
\label{sec4.2}

The purpose of this subsection is to show the emergence of the CF-type restrictions from the considerations of the 
(anti-)BRST and (anti-)co-BRST symmetry transformations of the Lagrangian densities~\eqref{1} and \eqref{10} which are
{\it different} from such kinds of considerations that have already been discussed in our 
Secs.~\ref{sec2} and \ref{sec3}. In this context, it is pertinent to point out that, for  the Lagrangian  
density~\eqref{1}, we have discussed {\it only} the off-shell nilpotent (co-)BRST symmetry transformations. However, 
this Lagrangian density respects the off-shell nilpotent anti-BRST as well as the anti-co-BRST symmetry  transformations, too, 
provided we take into account the validity of the CF-type restrictions~\eqref{22}. To corroborate this statement, 
we observe the following:
\begin{eqnarray}\label{26}
s_{ab} \, {\cal L}_{(B, {\cal B})} &=& - \partial_\mu \Big[\big(\partial^\mu \bar C^\nu 
- \partial^\nu \bar C^\mu \big) \bar B_\nu + \rho \big(\partial_\nu B^{\nu\mu} \big)  + B \, \partial ^\mu \bar C 
 + \dfrac{1}{2}\,\rho\, \bar B^\mu  - \dfrac{1}{2}\,\lambda \, \partial^\mu \bar \beta  \Big]\nonumber\\
&+& \big(\partial^\mu C^\nu - \partial^\nu C^\mu \big) \partial_\mu \Big[B_\nu + \bar B_\nu + \partial_\nu \phi \Big] 
+ \dfrac{1}{2} \big(\partial^\mu \rho \big) \Big[B_\mu + \bar B_\mu + \partial_\mu \phi \Big], 
\end{eqnarray} 
\begin{eqnarray}\label{27}
s_{ad} {\cal L}_{(B, {\cal B})} &=& \partial_\mu\Big[\varepsilon^{\mu\nu\sigma}\, \lambda\, \big(\partial_\nu A_\sigma \big) 
-  \big(\partial^\mu  C^\nu - \partial^\nu  C^\mu \big)\, {\cal B}_\nu 
+ {\cal B} \, \partial^\mu  C  + \dfrac{1}{2}\, \lambda\, \bar {\cal B}^\mu 
- \dfrac{1}{2}\, \rho \, \partial^\mu  \beta \Big]\nonumber\\
&+& \big(\partial^\mu  C^\nu - \partial^\nu  C^\mu \big) \partial_\mu \Big[{\cal B}_\nu + \bar {\cal B}_\nu 
+ \partial_\nu \widetilde \phi \Big] 
- \dfrac{1}{2}\,\big(\partial^\mu \lambda \big)\Big[{\cal B}_\mu 
+ \bar {\cal B}_\mu + \partial_\mu \widetilde \phi \Big]. 
\end{eqnarray}
A close and careful look at the above equations~\eqref{26} and \eqref{27} establishes that if we take 
into account the validity of the CF-type reinsertions: $B_\mu + \bar B_\mu + \partial_\mu \phi = 0, \;
{\cal B}_\mu + \bar {\cal B}_\mu + \partial_\mu \widetilde \phi = 0 $, we observe that the Lagrangian 
 density ${\cal L}_{(B, {\cal B})} $ [cf. Eq.~\eqref{1}],
not {\it only} respects the nilpotent (co-)BRST symmetry transformations [cf. Eqs.~\eqref{4},~\eqref{5}], it respects 
[cf. Eqs.~\eqref{26},~\eqref{27}] the nilpotent anti-BRST as well as the anti-co-BRST symmetry transformations, too. 
This is due to the fact that, on the submanifold of the fields where the CF-type restrictions~\eqref{22} are valid,  
the Lagrangian density~\eqref{1} respects {\it all} the fermionic symmetry transformations 
[cf. Eqs.~\eqref{2},~\eqref{3},~\eqref{11},~\eqref{12},~\eqref{25}] because it transforms to the total spacetime derivatives.

Against the backdrop of the above discussions, let us now focus on the Lagrangian density~\eqref{10} and 
its continuous symmetry transformations. We have already shown that the anti-BRST transformations~\eqref{11}
and anti-co-BRST  transformations~\eqref{12} are the {\it symmetry} transformations  for the Lagrangian density~\eqref{10}
because the {\it latter} transforms to the total spacetime derivatives [cf. Eqs.~\eqref{13},~\eqref{14}]
thereby rendering the action integral invariant (cf. Section~\ref{sec3}). We wish to demonstrate that the 
Lagrangian density~\eqref{10} {\it also} respects the (co-)BRST  transformations [cf. Eqs.~\eqref{2},~\eqref{3}] 
provided we take into account the validity of the CF-type restrictions~\eqref{22}. To corroborate this claim, we observe
that the following explicit transformations are true, namely;
\begin{eqnarray}\label{28}
s_d {\cal L}_{(\bar B, \bar {\cal B})} &=& - \,\partial_\mu\Big[\varepsilon^{\mu\nu\sigma}\, \rho\, \big(\partial_\nu A_\sigma \big)
 -  \big(\partial^\mu \bar C^\nu - \partial^\nu \bar C^\mu \big)\,\bar {\cal B}_\nu  - {\cal B} \, \partial^\mu  \bar C  
- \dfrac{1}{2}\, \rho\, {\cal B}^\mu - \dfrac{1}{2}\, \lambda \,\partial^\mu  \bar \beta  \Big]\nonumber\\
&-& \big(\partial^\mu \bar C^\nu - \partial^\nu \bar C^\mu \big)\, \partial_\mu \Big[{\cal B}_\nu 
+ \bar {\cal B}_\nu + \partial_\nu \widetilde \phi \Big] 
- \dfrac{1}{2}\,\big(\partial^\mu \rho \big)\Big[{\cal B}_\mu + \bar {\cal B}_\mu + \partial_\mu \widetilde \phi \Big], 
\end{eqnarray}
\begin{eqnarray}\label{29}
s_b \, {\cal L}_{(\bar B, \bar {\cal B})} &=& \partial_\mu \Big[\big(\partial^\mu C^\nu - \partial^\nu C^\mu \big) \bar B_\nu
+ \lambda \big(\partial_\nu B^{\nu\mu} \big) - B \,\partial ^\mu C  - \dfrac{1}{2}\,\lambda\, B^\mu  
+ \dfrac{1}{2}\,\rho\, \,\partial^\mu \beta  \Big]\nonumber\\
&-&  \big(\partial^\mu C^\nu - \partial^\nu C^\mu \big) \,\partial_\mu \Big[B_\nu + \bar B_\nu + \partial_\nu \phi \Big] 
+ \dfrac{1}{2} \big(\partial^\mu \lambda \big) \Big[B_\mu + \bar B_\mu + \partial_\mu \phi \Big],\qquad
\end{eqnarray}  
where the nilpotent co-BRST transformations ($s_d$) are listed in the equation~\eqref{2} and the nilpotent BRST 
transformations ($s_b$) are quoted in~\eqref{3}. It is straightforward to note that if we impose the CF-type restrictions: 
$B_\mu + \bar B_\mu + \partial_\mu \phi = 0, \, {\cal B}_\mu + \bar {\cal B}_\mu + \partial_\mu \widetilde \phi = 0$
on the above equation, we observe that, under the nilpotent (co-)BRST symmetry transformations, the 
Lagrangian density~\eqref{10} transforms to the following total spacetime derivatives:
\begin{eqnarray}\label{30}
s_d {\cal L}_{(\bar B, \bar {\cal B})} &=& - \,\partial_\mu\Big[\varepsilon^{\mu\nu\sigma}\, \rho\, \big(\partial_\nu A_\sigma \big)
 -  \big(\partial^\mu \bar C^\nu - \partial^\nu \bar C^\mu \big)\,\bar {\cal B}_\nu 
- {\cal B} \, \partial^\mu  \bar C  - \dfrac{1}{2}\, \rho\, {\cal B}^\mu 
- \dfrac{1}{2}\, \lambda \,\partial^\mu  \bar \beta  \Big], \nonumber\\
s_b \, {\cal L}_{(\bar B, \bar {\cal B})} &=& \partial_\mu \Big[\big(\partial^\mu C^\nu - \partial^\nu C^\mu \big) \bar B_\nu
+ \lambda \big(\partial_\nu B^{\nu\mu} \big) - B \,\partial ^\mu C  - \dfrac{1}{2}\,\lambda\, B^\mu  
+ \dfrac{1}{2}\,\rho\, \,\partial^\mu \beta  \Big].
\end{eqnarray} 
The above explicit expressions for the transformations of the Lagrangian density~\eqref{10} establish 
that the action integral $S = \int d^3 x\, {\cal L}_{(\bar B, \bar {\cal B})} $ remains invariant under 
the (co-)BRST symmetry transformations [cf. Eqs.~\eqref{2},~\eqref{3}] on the submanifold of fields where the 
CF-type restrictions~\eqref{22} are satisfied.

We wrap-up this subsection with the following concluding remarks. First of all, we lay emphasis on the 
fact that the existence of the {\it (non-)trivial} CF-type restrictions on a BRST-quantized gauge theory 
is as fundamental as the existence of the first-class constraints (see, e.g.~\cite{dir,sun,sm,git}) on the corresponding 
{\it classical} gauge theory. Second, the CF-type restrictions~\eqref{22} are unavoidable and in-built
{\it physical} restrictions on our $3D$ theory because they are (anti-)BRST as well as (anti-)co-BRST invariant 
quantities [cf. Eq.~\eqref{24}]. Third, these restrictions are very {\it robust} from the point of view of the 
symmetries because they {\it also} remain invariant under the discrete duality symmetry transformations~\eqref{6}
for the Lagrangian density~\eqref{1} and their analogues for the Lagrangian density~\eqref{10}. To be precise, 
the CF-type restrictions~\eqref{22} get exchanged with each-other under the {\it total} set of discrete duality 
symmetry transformations for the {\it coupled} Lagrangian densities~\eqref{1}) and \eqref{10}.  Fourth, both 
the Lagrangian densities~\eqref{1} and \eqref{10} respect all the four fermionic [i.e. (anti-)BRST and 
(anti-)co-BRST] symmetry transformations provided we take into account the validity of the CF-type restrictions~\eqref{22}. 
Fifth, both the Lagrangian densities~\eqref{1} and \eqref{10} are {\it coupled} and {\it equivalent} because 
of the existence of the CF-type restrictions~\eqref{22}. Sixth, we would like to point out that the off-shell 
nilpotent symmetry transformations~\eqref{25}, on the Nakanishi-Lautrup auxiliary fields, can be {\it also} 
derived from (i) the absolute anticommutativity requirements (i.e. $\{s_b, \; s_{ab} \} = 0, \; \{s_d, \; s_{ad} \} = 0 $) 
of the nilpotent (anti-)BRST and (anti-)co-BRST symmetry transformations on Lorentz vector (anti-)ghost fields: 
$(\bar C_\mu) C_\mu$, (ii) the invariance of the kinetic terms under the (anti-)BRST symmetry transformations, 
and (iii) the invariance of the gauge-fixing terms under the (anti-)co-BRST symmetry transformations 
(see, e.g. Appendix~\ref{secA} below for details). Finally, the analogue of the equation~\eqref{30} 
and ensuing discussions can be repeated for our observations in equations~\eqref{26} and \eqref{27}, too. 
However, for the sake of brevity, we have {\it not} written it explicitly.


\subsection{Absolute Anticommutativity Requirements on the Off-Shell Nilpotent Symmetries: CF-Type Restrictions}
\label{sec4.3}


Two of the sacrosanct properties of the (anti-)BRST and (anti-)co-BRST symmetry transformations of our 
$3D$ field-theoretic model are the off-shell nilpotency and absolute anticommutativity. The off-shell 
nilpotency property encodes the fermionic nature of these symmetries which is just like the $\mathcal{N} = 2$ 
SUSY transformations that transform a bosonic field to its fermionic counterpart and vice-versa. 
On the other hand, the absolute anticommutativity property between (i) the BRST and anti-BRST symmetry 
transformations (i.e. $\{s_b, \; s_{ab} \} = 0 $), and (ii) the co-BRST and anti-co-BRST symmetry 
transformations (i.e. $\{s_d, \; s_{ad} \} = 0 $) demonstrate the linear independence of these symmetry 
transformations. This property of linear independence, in the context of the BRST formalism, is 
completely {\it different} from the $\mathcal{N} = 2$ SUSY transformations which do {\it not} 
anticommute with each-other. As far as the off-shell nilpotent (anti-)BRST  symmetry transformations 
[that have listed in equations~\eqref{3},~\eqref{11} and \eqref{25}] are concerned, it is very interesting to
point out that we observe the following anticommutativity property, namely;
\begin{eqnarray}\label{31}
\big\{s_b, \; s_{ab} \big \} B_{\mu\nu} & = &-\,\partial_\mu \big[B_\nu + {\bar B}_\nu \big] +
 \partial_\nu \big[B_\mu + {\bar B}_\mu \big] \nonumber\\
&\equiv& - \,\partial_\mu \big[B_\nu + {\bar B}_\nu + \partial_\nu \phi \big] 
+  \partial_\nu \big[B_\mu + {\bar B}_\mu + \partial_\mu \phi \big]. 
 \end{eqnarray}
In other words, it is clear that the absolute anticommutativity property of the operator forms of the 
BRST and anti-BRST symmetry transformations  (i.e. $\{s_b, \; s_{ab} \} = 0 $) is obeyed if and only 
if we invoke the validity of the CF-type restriction: $B_\mu + \bar B_\mu + \partial_\mu \phi = 0$ [cf.~\eqref{22}]. 
It turns out that, for the {\it rest} of the fields of our $3D$ field-theoretic model (that is described by 
the Lagrangian densities~\eqref{1} and \eqref{10}), the sacrosanct property of absolute anticommutativity  
(i.e. $\{s_b, \; s_{ab} \} = 0 $) is {\it trivially} satisfied.

We concentrate now on the study of the absolute anticommutativity (i.e. $\{s_d, \; s_{ad} \} = 0$) 
between the off-shell nilpotent (i.e. $s_{(a)d}^2 = 0 $) co-BRST ($s_d$) and anti-co-BRST  ($s_{ad}$) 
symmetry transformations that have been quoted in the equations~\eqref{12},~\eqref{2} and \eqref{25}.
It is interesting to point out that the following operator form of the anticommutator, namely; 
\begin{eqnarray}\label{32}
\big\{s_d, \; s_{ad} \big \} A_\mu  &=& \varepsilon_{\mu\nu\sigma} \partial^\nu \big[{\cal B}^\sigma + \bar{\cal  B}^\sigma  \big] \nonumber\\
&\equiv&  \varepsilon_{\mu\nu\sigma} \partial^\nu \big[{\cal B}^\sigma + \bar{\cal  B}^\sigma 
+ \partial^\sigma \widetilde \phi \big],
 \end{eqnarray}
demonstrates that the absolute anticommutativity (i.e. $\{s_d, \; s_{ad} \} = 0$) of the off-shell nilpotent 
co-BRST and anti-co-BRST symmetry operators is satisfied if and only if we invoke the validity of the CF-type 
restriction: ${\cal B}_\mu + \bar {\cal B}_\mu + \partial_\mu \widetilde \phi = 0 $ [cf.~\eqref{22}]. We would like to lay 
emphasis on the fact that when the operator form of the above anticommutator (i.e. $\{s_d, \; s_{ad} \}$) acts on the
{\it rest} of the fields of the coupled (but equivalent) Lagrangian densities~\eqref{1} and \eqref{10} of our 
$3D$ field-theoretic model, it turns out to be {\it trivially} zero. This observation can be mathematically 
expressed, in a concise manner,  as:
\begin{eqnarray}\label{33}
&& \big\{s_d, \; s_{ad} \big \} \, \Phi = 0, \nonumber\\
&&  \Phi = B_{\mu\nu}, \, B_\mu, \bar B_\mu,\, {\cal  B}_\mu, \bar {\cal B}_\mu,\,
C_\mu, \, \bar C_\mu, \beta, \, \bar \beta, \, C, \, \bar C,\, \phi, \, \widetilde \phi, \, \rho, \, \lambda. 
 \end{eqnarray}
Thus, it is clear that we have to invoke the CF-type restriction: 
${\cal B}_\mu + \bar {\cal B}_\mu + \partial_\mu \widetilde \phi = 0 $ {\it only} in the proof of 
$ \big\{s_d, \; s_{ad} \big \} A_\mu = 0$ when the operator form of the anticommutator (i.e. $\{s_d, \; s_{ad} \}$)
acts on the Abelian 1-form gauge field $A_\mu$ of our $3D$ field-theoretic model.

We conclude this subsection with the following crucial remarks. First of all, we note that we have 
to invoke the CF-type restriction: $ B_\mu + \bar  B_\mu + \partial_\mu  \phi = 0 $ {\it only} in 
the proof of the absolute anticommutativity (i.e. $\{s_b, \; s_{ab} \} = 0$) between the BRST and 
anti-BRST symmetry transformations when the operator form of the anticommutator (i.e. $\{s_b, \; s_{ab} \}$) acts
on the Abelian 2-form gauge field $B_{\mu\nu}$ of our $3D$ theory. For the {\it rest} of the fields of our theory, 
it turns out that the absolute anticommutativity property (i.e. $\{s_b, \; s_{ab} \} = 0$) is {\it trivially} 
satisfied. Second, the absolute anticommutativity between the co-BRST and anti-co-BRST symmetry 
transformation operators is automatically satisfied for {\it all} the fields of our $3D$ theory {\it except} 
in the proof of $\big\{s_d, \; s_{ad} \big \} A_\mu = 0 $ [cf. Eq.~\eqref{32}] where we have to invoke the  
validity of the CF-type restriction: ${\cal B}_\mu + \bar {\cal B}_\mu + \partial_\mu \widetilde \phi = 0 $. 
Finally, it is obvious (in a subtle manner) that the requirements of the absolute anticommutativity properties 
between (i) the nilpotent BRST and anti-BRST symmetry transformations, and (ii) the nilpotent co-BRST and 
anti-co-BRST symmetry transformations, lead to the validity of the  {\it physical} 
CF-type restrictions~\eqref{22} on our $3D$ field-theoretic model. \\


\section{Bosonic Transformations: Uniqueness Property}
\label{sec5}


Out of the {\it four} infinitesimal, continuous and off-shell nilpotent (i.e. $s_{(a)b}^2 = 0, \; s_{(a)d}^2 = 0 $) 
(anti-)BRST ($s_{(a)b} $) and (anti-)co-BRST ($s_{(a)d} $) symmetry transformations, we have already seen 
(in the previous section) that the absolute anticommutativity property (i.e. $\{s_b, \; s_{ab} \} = 0$) 
between the off-shell nilpotent BRST and anti-BRST symmetry transformations is satisfied if and only if we invoke 
the validity of the CF-type resection: $ B_\mu + \bar  B_\mu + \partial_\mu  \phi = 0 $. On the other hand, 
the CF-type restriction: ${\cal B}_\mu + \bar {\cal B}_\mu + \partial_\mu \widetilde \phi = 0 $ has been
invoked in the proof of the absolute anticommutativity (i.e. $\big\{s_d, \; s_{ad} \big \} = 0 $) between 
the off-shell nilpotent co-BRST and anti-co-BRST symmetry transformations. In addition to these two 
anticommutativity properties, we have the following anticommutators 
\begin{eqnarray}\label{34}
\big\{s_d, \; s_{ab} \big \}  = 0, \quad \qquad \big\{s_b, \; s_{ad} \big \}  = 0,
 \end{eqnarray}
which are {\it trivially} satisfied. We are left with {\it two} more anticommutators which define the non-trivial 
infinitesimal and continuous bosonic (i. e. $s_\omega^2 \neq 0,\; s_{\bar \omega}^2 \neq 0 $) symmetry transformations 
($s_\omega, \; s_{\bar\omega} $) as:
\begin{eqnarray}\label{35}
\big\{s_d, \; s_{b} \big \}  = s_\omega, \quad\qquad \big\{s_{ad}, \; s_{ab} \big \}  = s_{\bar \omega}.
 \end{eqnarray}
The central theme of our present section is to show that only one of the above {\it two} bosonic symmetry 
transmissions is independent in the sense that the operator forms of these bosonic symmetry transformations 
satisfy: $s_\omega + s_{\bar \omega} = 0 $ on the submanifold of the quantum fields where the CF-type restrictions~\eqref{22} 
of our theory are satisfied.

To corroborate the above assertion, first of all, we note that we have the following 
\begin{eqnarray}\label{36}
&& s_\omega B_{\mu\nu} =  \big(\partial_\mu {\cal B}_\nu - \partial_\nu {\cal B}_\mu\big) 
+ \varepsilon_{\mu\nu\sigma}\, \big(\partial^\sigma B \big), 
\qquad  s_\omega A_\mu = \partial_\mu {\cal B} - \varepsilon_{\mu\nu\sigma}\, \big(\partial^\nu B^\sigma \big), \nonumber\\
&&   s_\omega C_\mu = \partial_\mu  \lambda, \qquad s_\omega \bar C_\mu = \partial_\mu \rho, \nonumber\\
&& s_\omega \big[B, \, {\cal B},\, \phi,\, \widetilde\phi,\, \beta,\, \bar \beta, \rho, \, 
\lambda, \, C, \,\bar C,\, B_\mu,\, \bar B_\mu,\, {\cal B}_\mu, \bar {\cal B}_\mu\big] = 0,\quad
\end{eqnarray}
\begin{eqnarray}\label{37}
&& s_{\bar \omega} B_{\mu\nu} = \big(\partial_\mu \bar {\cal B}_\nu - \partial_\nu \bar {\cal B}_\mu\big) 
 - \varepsilon_{\mu\nu\sigma}\, \big(\partial^\sigma B \big), \qquad
 s_{\bar \omega} A_\mu = - \partial_\mu {\cal B} - \varepsilon_{\mu\nu\sigma}\, \big(\partial^\nu \bar B^\sigma \big), \nonumber\\
&&  s_{\bar \omega} C_\mu = - \partial_\mu  \lambda, \qquad   s_{\bar \omega} \bar C_\mu = - \partial_\mu \rho, \nonumber\\
&& s_{\bar \omega} \big[B, \, {\cal B},\, \phi,\, \widetilde\phi,\, \beta,\, \bar \beta, \rho, \, 
\lambda, \, C, \,\bar C,\, B_\mu,\, \bar B_\mu,\, {\cal B}_\mu, \bar {\cal B}_\mu\big] = 0, \quad
\end{eqnarray} 
infinitesimal and continuous bosonic symmetry transformations for {\it all} the fields of our coupled 
(but equivalent) Lagrangian densities~\eqref{1} and \eqref{10}. A noteworthy point, at this stage, is 
the observation that the Faddeev-Popov (FP) (anti-)ghost fields of our $3D$ theory either do {\it not} 
transform at all under the bosonic symmetry transformations ($s_\omega, \; s_{\bar\omega} $) or they 
transform up to the $U(1)$ vector gauge symmetry-type transformations. It is, furthermore,  very interesting 
to mention here that, we observe the following expressions, namely;
\begin{eqnarray}\label{38}
\big(s_\omega + s_{\bar \omega} \big) \, B_{\mu\nu} &=& \partial_\mu\Big[{\cal B}_\nu + \bar {\cal B}_\nu \Big] 
- \partial_\nu\Big[{\cal B}_\mu + \bar {\cal B}_\mu \Big] \nonumber\\
&\equiv & \partial_\mu\Big[{\cal B}_\nu + \bar {\cal B}_\nu 
+ \partial_\nu \widetilde \phi \Big] - \partial_\nu\Big[{\cal B}_\mu + \bar {\cal B}_\mu 
+ \partial_\mu \widetilde \phi \Big], \nonumber\\
\big(s_\omega + s_{\bar \omega} \big) \, A_\mu &=& - \varepsilon_{\mu\nu\sigma}\, \partial^\nu\Big[B^\sigma 
+ \bar B^\sigma \Big] \nonumber\\
&\equiv& - \varepsilon_{\mu\nu\sigma}\, \partial^\nu\Big[B^\sigma 
+ \bar B^\sigma + \partial^\sigma \phi \Big],
\end{eqnarray}
which prove that the operator form of the sum (i.e. $s_\omega + s_{\bar \omega} = 0$) of the bosonic 
symmetry transformations (i.e. $s_\omega, \; s_{\bar\omega} $) turns out to be {\it zero} only on the 
submanifold of the  quantum fields of our $3D$ theory where the CF-type restrictions~\eqref{22} are satisfied. 
We observe that, for the {\it rest} of the fields of our theory, the operator form of the above sum turns 
out to be {\it trivially} zero as is evident from a close look at the transformations~\eqref{36} and \eqref{37}. 
Thus, it is clear that, for our {\it combined} system of the $3D$ free Abelian 1-form and 2-form gauge theories, the 
operator form of the sum $s_\omega + s_{\bar\omega} $ is {\it trivially} zero for all the fields {\it except} 
the Abelian 1-form and 2-form {\it basic} fields where the CF-type restrictions~\eqref{22} are required
to prove that: $s_\omega + s_{\bar \omega} = 0$ (as far as the bosonic symmetry transformations~\eqref{36} and \eqref{37} are concerned).

To prove that the infinitesimal and continuous bosonic  transformations~\eqref{36} and \eqref{37} are the 
{\it perfect} symmetry transformations for the coupled (but equivalent) Lagrangian densities~\eqref{1} and \eqref{10}, 
we note the following specific transformations
\begin{eqnarray}\label{39}
s_\omega {\cal L}_{(B, {\cal B})} &=& \partial_\mu \Big[\big(\partial^\mu {\cal B}^\nu - \partial^\nu {\cal B}^\mu \big) B_\nu 
- \big(\partial^\mu B^\nu - \partial^\nu B^\mu\big) {\cal B}_\nu - B \, \partial^\mu {\cal B} \nonumber\\
&+&  {\cal B} \,\partial^\mu B + \dfrac{1}{2}\, \lambda \, \partial^\mu \rho + \dfrac{1}{2}\, \rho \,\partial^\mu \lambda\Big],
\end{eqnarray}
\begin{eqnarray}
\label{40}
s_{\bar \omega} {\cal L}_{(\bar B, \bar {\cal B})} &=& \partial_\mu \Big[\big(\partial^\mu {\bar B}^\nu 
- \partial^\nu {\bar B}^\mu \big) \bar {\cal B}_\nu 
- \big(\partial^\mu \bar {\cal  B}^\nu - \partial^\nu \bar {\cal B}^\mu\big) {\bar B}_\nu + B \, \partial^\mu {\cal B} \nonumber\\
&-&  {\cal B} \, \partial^\mu B 
- \dfrac{1}{2}\, \lambda \, \partial^\mu \rho - \dfrac{1}{2}\, \rho \, \partial^\mu \lambda\Big],
\end{eqnarray}
which render the action integrals, corresponding to the Lagrangian densities~\eqref{1} and \eqref{10}, 
invariant due to Gauss's divergence theorem. It is interesting to point out that, as far as the 
transformations~\eqref{36} and \eqref{37} are concerned, we {\it also} observe the following 
\begin{eqnarray}\label{41}
s_\omega {\cal L}_{(\bar B, \bar {\cal B})} &=& \partial_\mu \Big[\big(\partial^\mu B^\nu - \partial^\nu B^\mu \big) \bar {\cal B}_\nu 
- \big(\partial^\mu  {\cal  B}^\nu - \partial^\nu {\cal B}^\mu\big) {\bar B}_\nu - B \, \partial^\mu {\cal B} \nonumber\\
&+&  {\cal B} \, \partial^\mu B + \dfrac{1}{2}\, \lambda \, \partial^\mu \rho
+ \dfrac{1}{2}\, \rho \, \partial^\mu \lambda \Big] \nonumber\\
&-& \big(\partial^\mu B^\nu - \partial^\nu B^\mu \big)\,\partial_\mu \Big[{\cal B}_\nu + \bar {\cal B}_\nu 
+ \partial_\nu \widetilde \phi \Big] \nonumber\\
&+& \big(\partial^\mu {\cal B}^\nu - \partial^\nu {\cal B}^\mu \big)\, \partial_\mu \Big[B_\nu + \bar B_\nu + \partial_\nu \phi \Big], 
\end{eqnarray}
\begin{eqnarray}\label{42}
s_{\bar \omega} {\cal L}_{(B, {\cal B})} &=& \partial_\mu \Big[\big(\partial^\mu \bar {\cal B}^\nu 
- \partial^\nu \bar {\cal B}^\mu \big) B_\nu - \big(\partial^\mu  {\bar  B}^\nu 
- \partial^\nu {\bar B}^\mu\big) {\cal B}_\nu + B \big(\partial^\mu {\cal B}\big) \nonumber\\
&-&  {\cal B} \, \partial^\mu B - \dfrac{1}{2}\, \lambda \big(\partial^\mu \rho\big)
- \dfrac{1}{2}\, \rho \big(\partial^\mu \lambda\big)\Big] \nonumber\\
&+& \big(\partial^\mu \bar B^\nu - \partial^\nu \bar B^\mu \big)\, \partial_\mu \Big[{\cal B}_\nu 
+ \bar {\cal B}_\nu + \partial_\nu \widetilde \phi \Big] \nonumber\\
&-& \big(\partial^\mu \bar {\cal B}^\nu - \partial^\nu \bar {\cal B}^\mu \big)\, 
\partial_\mu \Big[B_\nu + \bar B_\nu + \partial_\nu \phi \Big],  
\end{eqnarray}
which demonstrate that, if we invoke the validity of the CF-type restrictions~\eqref{22}, the coupled 
Lagrangian densities~\eqref{1} and \eqref{10} respect {\it both} the transformations~\eqref{36} and \eqref{37} 
{\it together}. Ultimately, it is worthwhile to mention that (i) the Lagrangian density~\eqref{1} respects the symmetry 
transformations~\eqref{36} in a {\it perfect} manner because we do {\it not} invoke any outside condition 
for the proof of \eqref{39}, (ii) the transformations~\eqref{37} are the {\it perfect} symmetry transformations 
[cf. Eq.~\eqref{40}] for the Lagrangian  density~\eqref{10} (in exactly the same manner as (i)), and (iii) the 
transformations~\eqref{36} and \eqref{37} are the symmetry transformations for the Lagrangian 
densities~\eqref{10} and \eqref{1}, respectively, provided  we use the sanctity of the CF-type 
restrictions~\eqref{22} on the r.h.s. of the transformations~\eqref{41} and \eqref{42}. 
It is worthwhile to point out that there is a simpler method to derive the transformations 
\eqref{39} and \eqref{40} where we can use $s_\omega {\cal L}_{(B, {\cal B})}  = \{s_b, \; s_d \} \,{\cal L}_{(B, {\cal B})}$, 
~$s_{\bar \omega} {\cal L}_{(\bar B, \bar {\cal B})}  = \{s_{ab}, \; s_{ad} \} \,{\cal L}_{(\bar B, \bar{\cal B})} $
and exploit our earlier results of Secs.~\ref{sec2} and~\ref{sec3} where we have already obtained the transformations
\eqref{4},~\eqref{5},~\eqref{13} and~\eqref{14}. In exactly similar fashion, we  can  utilize the beauty
of the equations [cf. Eqs.~\eqref{26},~\eqref{27},~\eqref{28} and \eqref{29}]  to prove the validity 
of~\eqref{41} and \eqref{42}. All these results and inputs can be further utilized in
 the proofs of equations~\eqref{43} and \eqref{44}.

Before we end this short section, we would like to point out that the operator equation:  
$s_\omega + s_{\bar \omega} = 0$  is {\it also} true at the level of  symmetry transformations 
of the Lagrangian densities~\eqref{1} and \eqref{10} provided we use the validity of the CF-type 
restrictions~\eqref{22} because we note that the following explicit transformations are true:
\begin{eqnarray}\label{43}
\big(s_\omega + s_{\bar \omega} \big) \,{\cal L}_{(B, {\cal B})} &=& \partial_\mu\Big[\partial^\mu \big({\cal B}^\nu 
+ \bar {\cal B}^\nu + \partial^\nu \widetilde \phi \big) B_\nu 
- \partial^\nu \big({\cal B}^\mu + \bar {\cal B}^\mu + \partial^\mu \widetilde \phi \big) B_\nu \nonumber\\
&-& \partial^\mu \big(B^\nu + \bar B^\nu + \partial^\nu \phi \big) {\cal B}_\nu 
+ \partial^\nu \big(B^\mu + \bar B^\mu + \partial^\mu \phi \big) {\cal B}_\nu \Big] \nonumber\\
&+&\big(\partial^\mu \bar B^\nu - \partial^\nu \bar B^\mu \big)\partial_\mu \Big[{\cal B}_\nu 
+ \bar {\cal B}_\nu + \partial_\nu \widetilde \phi \Big] \nonumber\\
&-& \big(\partial^\mu \bar {\cal B}^\nu - \partial^\nu \bar {\cal B}^\mu \big)\,\partial_\mu \Big[B_\nu 
+ \bar B_\nu + \partial_\nu \phi \Big],
\end{eqnarray}
\begin{eqnarray}\label{44}
\big(s_\omega + s_{\bar \omega} \big) {\cal L}_{(\bar B, \bar{\cal B})} &=& \partial_\mu\Big[\partial^\mu \big(B^\nu 
+ \bar B^\nu + \partial^\nu \phi \big) \bar {\cal B}_\nu 
- \partial^\nu \big(B^\mu + \bar B^\mu + \partial^\mu \phi \big) \bar {\cal B}_\nu \nonumber\\
&-& \partial^\mu \big({\cal B}^\nu + \bar {\cal B}^\nu + \partial^\nu \widetilde \phi \big) {\bar B}_\nu 
+ \partial^\nu \big({\cal B}^\mu + \bar {\cal B}^\mu + \partial^\mu \widetilde \phi \big) {\bar B}_\nu \Big] \nonumber\\
&-&\big(\partial^\mu B^\nu - \partial^\nu B^\mu \big)\, \partial_\mu \Big[{\cal B}_\nu + \bar {\cal B}_\nu 
+ \partial_\nu \widetilde \phi \Big] \nonumber\\
&+& \big(\partial^\mu {\cal B}^\nu - \partial^\nu {\cal B}^\mu \big)\, \partial_\mu \Big[B_\nu + \bar B_\nu + \partial_\nu \phi \Big].
\end{eqnarray}
In other words, we observe that the sum of the infinitesimal and continuous bosonic symmetry transformations 
(i.e. $s_\omega + s_{\bar \omega}$) turns out to be {\it zero} when we apply this specific combination
on the {\it coupled} Lagrangian densities~\eqref{1} and \eqref{10} provided we use the sanctity of the
CF-Type restrictions: $B_\mu + \bar B_\mu + \partial_\mu \phi = 0, \, 
{\cal B}_\mu + \bar {\cal B}_\mu + \partial_\mu \widetilde \phi = 0$ [cf. Eq.~\eqref{22}]. Thus, it is crystal clear that only 
{\it one} of the two bosonic symmetry transformations~\eqref{36} and \eqref{37} is {\it independent} (i) at 
the level of the {\it basic} Abelian 2-form and 1-form fields [cf. Eq.~\eqref{38}] of the  Lagrangian 
densities~\eqref{1} and \eqref{10}, and (ii) at the level of the symmetry invariance [cf. Eqs.~\eqref{43},~\eqref{44}]
of the {\it coupled} Lagrangian densities~\eqref{1} and \eqref{10}.

We conclude this section with the following {\it final} remarks. First of all, we observe that the (anti-)ghost 
fields of our theory either do {\it not} transform at all or they transform up to the $U(1)$ gauge symmetry-type 
transformations under the bosonic symmetry  transformations [cf. Eqs.~\eqref{36},~\eqref{37}]. Second, at the operator level, 
only {\it one} of the bosonic symmetry transformations~\eqref{36} and \eqref{37} is {\it independent}  on the submanifold of 
quantum fields where the CF-type restrictions~\eqref{22} are satisfied. This proves the {\it uniqueness} of the bosonic 
symmetry transformation operator because we observe that 
$s_\omega + s_{\bar \omega} = 0$ due to the validity of physical restrictions in Eq.~\eqref{22}. Third, we note that the {\it coupled} Lagrangian densities~\eqref{1} and \eqref{10} respect {\it both} 
the bosonic symmetry transformations~\eqref{36} and \eqref{37} {\it together} [cf. Eqs.~\eqref{39},~\eqref{40},~\eqref{41},~\eqref{42}] 
provided we take into account the validity of the  CF-type restrictions~\eqref{22}. Fourth, the {\it unique} bosonic symmetry 
operator commutes with all the fermionic (i.e. off-shell nilpotent) (anti-)BRST and (anti-)co-BRST symmetry 
transformation operators. This property can be proven in a straightforward manner.  If we take $s_\omega = s_b \,s_d + s_d \,s_b $ 
to be the independent and unique bosonic symmetry operator, it is straightforward to note that 
\begin{eqnarray}\label{45}
\big [s_\omega, \; s_{b} \big ]  =  s_b\, s_d \, s_b -  s_b\, s_d \, s_b = 0, \nonumber\\
\big [s_\omega, \; s_{d} \big ]  =  s_d\, s_b \, s_d -  s_d\, s_b \, s_d = 0,
 \end{eqnarray}
where we have to take into account the off-shell nilpotency properties (i.e. $s_b^2 = 0, \; s_d^2 = 0 $) 
of the BRST ($s_b$) and co-BRST ($s_d$) symmetry transformation operators. If we take into account 
the validity of the operator equation: $s_\omega + s_{\bar \omega} = 0$ (which proves the {\it uniqueness} 
of the bosonic symmetry operator), it is pretty easy to prove that the following are true, namely;
\begin{eqnarray}\label{46}
\big [s_\omega, \, s_{ab} \big ] &\equiv & - \, \big [s_{\bar \omega}, \, s_{ab} \big ]  
= s_{ab}\, s_{ad} \, s_{ab} -  s_{ab} \,s_{ad} \, s_{ab}  = 0, \nonumber\\
\big [s_\omega, \, s_{ad} \big ]  &\equiv & - \, \big [s_{\bar \omega}, \, s_{ad} \big ] 
=  s_{ad}\, s_{ab}\,  s_{ad} -  s_{ad}\, s_{ab} \, s_{ad}  = 0,
 \end{eqnarray}
where we have taken into account (i) the validity of the off-shell nilpotency property 
(i.e. $s_{ab}^2 = 0, \; s_{ad}^2 = 0 $) of the anti-BRST ($s_{ab}$) and anti-co-BRST ($s_{ad}$) 
symmetry operators, (ii) the definition of the bosonic symmetry transformation operator: 
$s_{\bar\omega} = s_{ab} \,s_{ad} + s_{ad} \,s_{ab} $, and (iii) the validity of the CF-type 
restrictions~\eqref{22} which play a crucial role in the proof of the operator equation: 
$s_\omega + s_{\bar \omega} = 0$. Finally, the observations, made in equations~\eqref{45} and \eqref{46}, 
play a crucial role in the discussion on the algebraic structures that are obeyed by the
symmetry transformation operators of our theory (cf. Section~\ref{sec7} below) where we 
demonstrate that the unique bosonic symmetry operator commutes with all the 
 {\it six} continuous symmetry transformation operators of our $3D$ BRST-quantized field-theoretic system.


\section{Ghost-Scale Symmetry Transformations}
\label{sec6}

In the coupled (but equivalent) Lagrangian densities~\eqref{1} and \eqref{10}, the FP-ghost parts are exactly the 
{\it same}. On the other hand, the non-ghost (i.e physical) parts of the above Lagrangian densities are 
{\it different} due to (i) the use of different kinds of symbols for the Nakanishi-Lautrup auxiliary  fields, and 
(ii) different signs for the (pseudo-)scalar fields $(\widetilde \phi)\phi$. However, the non-ghost parts of the 
Lagrangian densities are {\it equivalent} on the submanifold of the quantum fields where the CF-type restrictions~\eqref{22} 
are satisfied. It is interesting to point  out that under the following ghost-scale symmetry transformations
\begin{eqnarray}\label{47}
&& C_\mu \to e^{+\,\Omega}\, C_\mu,  \qquad \bar C_\mu \to e^{-\,\Omega}\, \bar C_\mu, \qquad
 \beta \to e^{+2\,\Omega}\, \beta, \qquad  \bar \beta \to e^{-2\,\Omega}\, \bar \beta, \nonumber\\
&& C \to e^{+\,\Omega}\, C, \qquad\;\;\;  \bar C \to e^{-\,\Omega}\, \bar C, 
\qquad \;\;\;\lambda \to e^{+ \,\Omega}\, \lambda, \qquad\;\;  \rho \to e^{- \,\Omega}\, \rho, \nonumber\\
&& \Psi \to e^{ 0\,\Omega}\,\Psi, \qquad\;\;\; \Psi = B_{\mu\nu},\, A_\mu,\, B_\mu, \,
\bar B_\mu,\, {\cal B}_\mu, \, \bar {\cal B}_\mu, \, B, \,{\cal B}, \phi, \, \widetilde \phi,
\end{eqnarray}
the Lagrangian densities~\eqref{1} and \eqref{10} remain invariant. In the above transformations, the transformation 
parameter $\Omega$ is a global  (i.e. spacetime independent) scale parameter and the numerals in the exponents 
denote the ghost numbers of the fields. It is now crystal clear that (i) the physical (i.e. non-ghost) generic 
field $\Psi$ has the ghost number equal to zero, and (ii) the fermionic (i.e. $\rho\, \lambda + \lambda\, \rho = 0 $)
auxiliary fields $(\rho)\lambda$ carry the ghost numbers $(-1)+1$, respectively,  because of the relationships: 
$\rho = - 2\, (\partial \cdot \bar C) $ and $\lambda = + 2\,(\partial \cdot  C)$ which emerge out as the EL-EoMs 
from the Lagrangian densities~\eqref{1} and/or  \eqref{10}. For the sake of brevity, we set the global scale parameter  
$\Omega = 1$. This leads to the derivation of the infinitesimal version ($s_g$) of the ghost-scale transformations~\eqref{47} as 
\begin{eqnarray}\label{48}
&& s_g C_\mu = + \,C_\mu, \qquad  s_g \bar C_\mu = - \, \bar C_\mu, 
\qquad  s_g \beta = + 2 \, \beta, \qquad  s_g \bar \beta = - 2 \, \bar \beta, \nonumber\\
&& s_g C = + \, C,  \qquad s_g \bar C = - \, \bar C, 
\qquad s_g \lambda = + \, \lambda, \qquad s_g \rho = -  \, \rho, \nonumber\\
&& s_g \Psi \equiv s_g [B_{\mu\nu}, A_\mu, B_\mu,  \bar B_\mu, {\cal B}_\mu,  
\bar {\cal B}_\mu,  B, {\cal B}, \phi,  \widetilde \phi ] = 0,  \quad\;\;\;\;
\end{eqnarray}
which leave the {\it coupled} Lagrangian densities~\eqref{1} and \eqref{10} perfectly invariant 
(i.e. $s_g {\cal L}_{(B, {\cal B})} = 0, \; s_g {\cal L}_{(\bar B, \bar{\cal B})} = 0$) and, hence, the 
corresponding action integrals, too.

We note that the operator form of the infinitesimal version ($s_g$) of the ghost-scale symmetry transformations 
(with $\Omega = 1$) [cf. Eq.~\eqref{48}] respects the following operator algebra with the {\it rest} of the symmetry 
operators  $ s_r$ (with $ r = b,\, ab,\, d,\,ad,\, \omega, \, g $), namely;
\begin{eqnarray}\label{49}
&&  \big[s_g, \,s_b \big] =+  s_b, \qquad  \big[s_g, \,s_{ab} \big] = -  s_{ab}, 
\qquad \big[s_g, \, s_d \big] = -  s_d, \nonumber\\
&&  \big[s_g, \,s_{ad} \big] = + s_{ad},  \qquad \big[s_g, \, s_{\omega} \big] = 0, 
\qquad \big[s_g, \;s_{g} \big] = 0,
\end{eqnarray}
where we have taken into account all the operator forms of the symmetry transformations that have been quoted in 
equations~\eqref{2},~\eqref{3},~\eqref{11},~\eqref{12},~\eqref{25}, \eqref{36} and \eqref{48}. Physically, the 
algebraic structures in~\eqref{49} imply the fact that (i) the BRST and anti-co-BRST symmetry transformations 
[cf. Eqs.~\eqref{3},~\eqref{12}] raise the ghost number of a field by {\it one} on which they act, (ii) the ghost 
number of a field, on the other hand,  is lowered by {\it one} when it is operated upon by the anti-BRST and co-BRST 
symmetry  transformations [cf. Eqs.~\eqref{11},~\eqref{2}], and (iii) the ghost number of a field remains intact when 
it is acted upon by the bosonic symmetry operator [cf. Eq.~\eqref{36}] and the ghost symmetry operator [cf. Eq.~\eqref{48}]. 
These observations will play a key role in our later discussions on the complete algebraic structures and their 
relationship with the cohomological operators (cf. Section~\ref{sec7} below).

We conclude this short section with the following useful remarks. First, we have seen that the {\it non-trivial} 
ghost-scale symmetry transformations [cf. Eqs.~\eqref{47},~\eqref{48}]
are confined {\it only} to the (anti-)ghost and fermionic auxiliary  [i.e.  $(\rho)\lambda$] 
fields in the FP-ghost parts of the coupled (but equivalent) Lagrangian densities~\eqref{1} and \eqref{10}. Second, 
the physical (i.e. non-ghost) fields, with ghost number equal to zero, transform {\it trivially} under the 
ghost-scale symmetry transformations [cf. Eqs.~\eqref{47},~\eqref{48}]. Third, all the {\it four} fermionic 
(i.e. off-shell nilpotent) (anti)BRST and (anti-)co-BRST symmetry transformations lead to the transformations of 
the bosonic fields to fermionic fields and vice-versa. Hence, the ghost numbers of the transformed fields 
{\it change} (within the framework of BRST formalism). The ghost numbers of the transformed fields are determined 
by the arguments that have been made after the algebraic relationships~\eqref{49}. Fourth, the bosonic and ghost-scale
symmetry transformations [cf. Eqs.~\eqref{36},~\eqref{48}] do {\it not} change the ghost number(s) of the field(s) 
on which they operate. In other words,  under the bosonic and ghost-scale symmetry transformations, the bosonic 
fields transform to the bosonic fields and the fermionic fields {\it obviously}  transform to the fermionic fields. 
Finally, the algebraic relationships in~\eqref{49} have connections with some of the key properties 
 of  the de Rham cohomological operators  of differential geometry (e.g. their operations on a form)  
which we shall discuss in the next section where we shall see that the degree of a given form will
be identified with the ghost number of a {\it specific} field that is present in the coupled (but equivalent) 
Lagrangian densities of  our $3D$ field-theoretic model of the BRST-quantized theory.\\


\section{Algebraic Structures: Cohomological Operators}
\label{sec7}

The central purpose of this section is to establish a deep relationship between (i) the algebraic structures that are 
obeyed by the discrete and continuous symmetry transformation operators of our BRST-quantized $3D$ {\it combined} field-theoretic system of 
the free Abelian 1-form and 2-form gauge theories, and (ii) the Hodge algebra that is respected by the de Rham 
cohomological operators of differential  geometry~\cite{egh,mm,gss,van,nis,joel}. To this end in our mind, first of all, 
we collect the complete set of algebraic structures that has been discussed so far.  In other words, we observe that 
the following algebra is satisfied by the transformation operators of our theory, namely;
\begin{eqnarray}\label{50}
&& s^2_b = 0, \qquad  s^2_{ab} = 0, \qquad s^2_d = 0, \qquad  s^2_{ad} = 0, 
\qquad s_\omega = \big\{s_d, \,s_b \big\} \equiv - s_{\bar\omega}, \nonumber\\
&& \big\{s_b,\, s_{ab} \big\} = 0, \qquad \big\{s_d,\, s_{ad} \big\} = 0, 
\qquad \big\{s_b, s_{ad} \big\} = 0, \qquad \big\{s_d, \,s_{ab} \big\} = 0, \nonumber\\
&& \big[s_g, \,s_b \big] = + s_b, \qquad \big[s_g, \,s_{ab} \big] = - s_{ab}, 
\qquad \big[s_g, \,s_d \big] = - s_d, \qquad  \big[s_g, \,s_{ad} \big] = + s_{ad},  \nonumber\\
&& s_{(a)d} = \pm *\,s_{(a)b} \,*, \quad\qquad s_{(a)b} = \mp * \,s_{(a)d}\, *,  \nonumber\\
&&  \big[s_\omega, \, s_r \big] = 0, \quad\qquad r= b,\, ab,\, d,\,ad,\, g, \, \omega,
\end{eqnarray}
for our $3D$ BRST-quantized field-theoretic model of a {\it combined} system of the free Abelian 1-form and 2-form gauge 
theories. A close and careful look at the above algebra reveals that the {\it unique} infinitesimal bosonic symmetry 
transformation operator ($s_\omega $) commutes with all the {\it rest} of the symmetry transformation operators of 
our theory.  On the other hand, the infinitesimal and continuous ghost-scale symmetry transformation operator has a  specific
kind of the algebraic structures with the continuous (anti-) BRST, (anti-)co-BRST, bosonic and ghost-scale  symmetry transformation 
operators which have been discussed in the previous section [cf. Eq.~\eqref{49}]. Lastly, as  discussed in Section~\ref{sec4}, 
the absolute anticommutativity properties (i.e. $\{ s_b, \, s_{ab} \} = 0, \, \{ s_d, \, s_{ad} \} = 0$) between (i) 
the BRST and anti-BRST  transformation operators [cf. Eq.~\eqref{31}], (ii) the co-BRST and anti-co-BRST  transformation 
operators [cf. Eq.~\eqref{32}], and (iii) the validity of the operator equation: $s_\omega + s_{\bar\omega} = 0 $ for 
the bosonic symmetry transformations [cf. Eqs.~\eqref{36},~\eqref{37},~\eqref{38}], are satisfied if and only if we invoke 
the validity [cf. Eqs.~\eqref{31},~\eqref{32},~\eqref{38}] of the CF-type restrictions~\eqref{22}.

It is very interesting to mention that the above algebraic structures [cf. Eq.~\eqref{50}] are reminiscent of the 
following algebraic structures that are obeyed  by the well-known set of {\it three} de Rham cohomological operators of differential 
geometry~\cite{egh,mm,gss,van,nis,joel}, namely;  
\begin{eqnarray}\label{51}
&& d^2 = 0, \qquad \delta^2 = 0, \qquad \{d, \; \delta\} =
\Delta=  \big(d + \delta \big)^2, \nonumber\\
&& \big[\Delta, \;  d \big] = 0, \qquad  \big[\Delta, \; \delta \big] = 0, \qquad \delta = \pm \, *\, d \, *. \qquad 
\end{eqnarray}
A close look at the equations~\eqref{50} and \eqref{51} and their precise comparison establish that there is a 
two-to-one mapping between the infinitesimal symmetry transformation operators  and the de Rham cohomological 
operators of differential geometry as follows:
\begin{eqnarray}\label{52}
&& \big(s_b, \; s_{ad} \big) \Longrightarrow d, \qquad  \quad\big(s_{ab}, \; s_{ad}\big) \Longrightarrow \delta, \nonumber\\
&& \big(s_\omega, \; s_{\bar \omega} \equiv -\,s_\omega \big)  \Longrightarrow \Delta. 
\end{eqnarray}
We further note that the algebraic relationship ($\delta = \pm \, *\, d \, *  $) between the nilpotent 
(i.e. $d^2 = 0, \; \delta^2 = 0$) (co-)exterior derivatives $(\delta)d$ of differential geometry is realized 
(i.e. $ s_{(a)d} = \pm *\,s_{(a)b} \,*, \; s_{(a)b} = \mp * \,s_{(a)d}\, * $) in terms of the interplay 
[cf. Eqs.~\eqref{7},~\eqref{9},~\eqref{15}]  between the infinitesimal and continuous  nilpotent 
(i.e. $s_{(a)b}^2 = 0, \,s_{(a)d}^2 = 0  $) (anti-)BRST ($s_{(a)b}$) and (anti-)co-BRST ($s_{(a)d}$)  
symmetry transformations [cf. Eqs.~\eqref{2},~\eqref{3},~\eqref{11},~\eqref{12},~\eqref{25}]  and the 
discrete duality symmetry transformations [cf. Eq.~\eqref{6}].  
The {\it latter} transformations provide the physical realization of the Hodge duality operator.

Before we wrap-up this section, we would like to point out some of the key properties (associated with  the 
cohomological operators) that are connected with their operations on a given well-defined form $f_{n}$ of degree 
$n$. We note that, when the exterior derivative $d$ acts on it, the degree of the ensuing form is raised by 
one (i.e. $d \,f_{n} \sim f_{n + 1}$). On the other hand, when $f_{n}$ is operated upon by the co-exterior
derivative $\delta$, the degree of the resulting form is lowered by one (i.e. $\delta\, f_{n} \sim f_{n - 1} $). 
Finally, the degree of a given form $f_{n}$ remains intact when it is acted upon by the Laplacian operator 
(i.e. $\Delta\, f_{n} \sim f_{n} $). These properties can be captured in the language of the symmetry transformation 
operators of our BRST-quantized $3D$ field-theoretic model. We have already commented on the ghost number(s) of the 
transformed field(s) under the nilpotent (anti-)BRST, (anti-)co-BRST and {\it unique} bosonic symmetry transformations
after equation~\eqref{49} of our previous section. Within the framework of BRST formalism, the degree of a given form 
(in the context of the differential geometry) can be identified with the ghost number of the specific field on 
which the {\it above} symmetry transformation operators act. As a result of this identification, we have been able 
to obtain a two-to-one mapping between the symmetry transformation operators and the de Rham cohomological  
operators of the differential geometry at the algebraic level [cf. Eq.~\eqref{52}].

We end  this section with the following remarks. First of all, we note that the infinitesimal, continuous 
and off-shell nilpotent (anti-)BRST and (anti-)co-BRST symmetry transformation operators 
[cf. Eqs.~\eqref{2},~\eqref{3},~\eqref{11},~\eqref{12},~\eqref{25}] are identified with the (co-)exterior 
derivatives of differential geometry and discrete duality symmetry transformations~\eqref{6} provide the physical 
realizations of the Hodge duality $*$ operator [cf. Eqs.~\eqref{7},~\eqref{9},~\eqref{15},~\eqref{50}]. 
In other words, it is the interplay between the off-shell nilpotent continuous symmetry transformation 
operators and the discrete duality symmetry transformation operator that provide the physical realizations 
of the mathematical relationship (i.e. $\delta = \pm \, *\, d\, * $) that exists between the (co-)exterior derivatives
of differential geometry~\cite{egh,mm,gss,van,nis,joel}.  Second, the change in the degree of a given 
differential form due to the operations of the cohomological operators is physically realized in terms the 
change in the ghost number of a specific field due to the operations of the (anti-)BRST, (anti-)co-BRST and 
{\it unique} bosonic symmetry transformation operators of our BRST-quantized $3D$ field theoretic model. 
Finally, at the algebraic level, the symmetry transformation operators of our $3D$ BRST-quantized theory 
and de Rham cohomological operators of differential geometry are identical in the sense that there exist 
a two-to-one mapping [cf. Eq.~\eqref{52}] between them. Hence, our $3D$ BRST-quantized field-theoretic 
model provides an example for Hodge theory.\\


\section{Conclusions}
\label{sec8}

In our present endeavor, we have concentrated on the symmetry properties of a $3D$ combined field-theoretic system of 
the free Abelian 1-form and 2-form gauge theories which is described by the coupled (but equivalent) Lagrangian densities 
[cf. Eqs.~\eqref{1},~\eqref{10}] that respect a set of {\it six} continuous symmetry transformations and a couple of very 
useful discrete duality symmetry transformations~\eqref{6}. These symmetries, in their operator form, obey an extended 
BRST algebra~\eqref{50}  which is reminiscent of the Hodge algebra [cf. Eq.~\eqref{51}] that is satisfied by the de Rham 
cohomological operators of differential geometry~\cite{egh,mm,gss,van,nis,joel}. To be precise, the extended BRST algebra 
contains more information than the Hodge algebra~\eqref{51} because of the presence of the hidden algebraic structure that 
incorporates the ghost-scale symmetry transformation operator [cf. Eq.~\eqref{49}]. Thus, the algebraic 
structures~\eqref{49} as well as~\eqref{51} are hidden in the extended BRST algebra~\eqref{50}. We have established that 
the off-shell nilpotent (i.e. $s_{(a)b}^2 = 0, \;s_{(a)d}^2 = 0 $), infinitesimal and continuous (anti-)BRST ($s_{(a)b}$) and 
(anti-)co-BRST ($s_{(a)d} $) transformation operators provide the physical realization(s)  of the nilpotent 
(i.e. $d^2 = 0, \;\delta^2 = 0$) (co-)exterior derivatives $(\delta)d$ of  differential geometry~\cite{egh,mm,gss,van,nis,joel}. 
On the other hand, the discrete duality transformation operators~\eqref{6} lead to the physical realization of the Hodge duality 
$*$ operation of differential geometry in the relationship: $\delta = \pm *\,d \, *$ between 
the nilpotent ($\delta^2 = 0, \; d^2 = 0 $) (co-)exterior derivatives $(\delta)d$.

Against the backdrop of the above paragraph, as a side remark, we would like to point  out that,  under the 
discrete duality symmetry transformations~\eqref{6}, we observe that (i) the total Lagrangian density~\eqref{1} 
remains invariant, (ii) in the ghost-sector, there is exchange: 
$-\, \frac{1}{2}\, (\partial \cdot C - \frac{1}{4}\, \lambda)\, \rho \leftrightarrow
 -\, \frac{1}{2}\, (\partial \cdot \bar C + \frac{1}{4}\, \rho)\, \lambda $ between {\it these} two terms and other terms 
remain invariant on their own, and  (iii) in the non-ghost sector $(a)$ the kinetic term of the Abelian 1-form gauge field 
exchanges with the gauge-fixing term of the  Abelian 2-form gauge field, and $(b)$ the kinetic term of the Abelian 2-form 
gauge field exchanges with the gauge-fixing term of the  Abelian 1-form gauge field (for our $3D$ BRST-quantized theory).

We would like to dwell a bit on the relationships: $s_{(a)d} = \pm *\,s_{(a)b} \,*, \;s_{(a)b} = \mp * \,s_{(a)d}\, *$ 
[cf. Eq.~\eqref{50}] in terms of the symmetry transformation operators that provide the physical realizations of the 
mathematical relationship: $\delta = \pm\, *\, d\, * $ with emphasis on the infinitesimal ghost-scale symmetry 
transformation operator ($s_g $) that is present in the algebraic structures~\eqref{50}.  In particular, the algebraic 
structures with the ghost-scale symmetry transformation operator [cf. Eqs.~\eqref{49},~\eqref{50}] establish that the 
pair of nilpotent symmetry transformation operators ($s_b, \; s_{ad} $) raise the ghost number of a field by one 
[cf.  Eqs.~\eqref{3},~\eqref{12}] on which they act. On the contrary, the operations of the other pair of nilpotent 
symmetry transformation operators ($s_d, \; s_{ab} $) lead to the lowering of the ghost number of a field by one 
[cf. Eqs.~\eqref{2},~\eqref{11}] on which they operate directly. These observations are 
analogous to the operations of the  exterior and co-exterior derivatives of differential geometry on a given differential form because 
we know that the exterior derivative raises the degree of a form by one and the co-exterior derivative lowers the 
degree of a form by one on which they operate. Hence, in the true sense of the similarity and
identification, the precise physical realizations of the relationship:  $\delta = \pm\, *\, d\, * $, in terms
of the symmetry transformation operators,  are: $s_{d} = \pm *\,s_{(b} \,*, \;s_{ab} = \mp * \,s_{ad}\, *$ which are 
picked out from the {\it total} relationships: $s_{(a)d} = \pm *\,s_{(a)b} \,*, \;s_{(a)b} = \mp * \,s_{(a)d}\, *$ 
on the basis of the identification of the ghost number of a field (that is present in our $3D$ BRST-quantized theory) 
with the degree of a differential form.

We have laid quite a bit of emphasis on the existence of CF-type restrictions [cf. Eq.~\eqref{22}] on our 
BRST-quantized $3D$ field-theoretic model and derived them from different theoretical angles 
(cf. Section~\ref{sec4} for details). In these derivations, mostly the fermionic (i.e. off-shell nilpotent) 
symmetry transformations have been given utmost importance.  We would like to point out that the CF-type 
restrictions~\eqref{22} {\it also} play very important roles in the proof of  the {\it uniqueness} 
of the bosonic symmetry transformation  operators (cf. Section~\ref{sec5} for details). In particular, we would like
to pinpoint that, in proof of the {\it uniqueness} of the bosonic symmetry transformation operator, its role
becomes very clear when we concentrate on the sum of their operations on the {\it basic} Abelian 1-form and 
2-from gauge fields of our theory [cf. Eq.~\eqref{38}]. To be precise, the CF-type restrictions appear 
in our equations \eqref{41},~\eqref{42},~\eqref{43} and~\eqref{44} where we have considered the operations 
of the individual as well as  the sum of bosonic symmetry transformation operators on the coupled
(but equivalent) Lagrangian densities~\eqref{1} and~\eqref{10} of our $3D$ BRST-quantized theory.

It is very interesting to point out that the (pseudo-) scalar fields $(\widetilde \phi)\phi$ are present in our 
BRST-quantized theory which is described by the Lagrangian densities~\eqref{1} and \eqref{10}. Both these fields 
are massless fields because they satisfy the {\it massless} Klein-Gordon equations of motion: 
$\Box \widetilde \phi = 0,\; \Box \phi = 0 $. However, there is a key {\it difference} between the two in the sense that
the scalar field ($\phi $) carries the {\it positive} kinetic term in contrast to the pseudo-scaler field 
($\widetilde \phi $) which is endowed with the {\it negative} kinetic term. Such fields, with the negative kinetic terms, 
have become quite popular in the realm of the cyclic, bouncing and self-accelerated 
cosmological models of the Universe (see, e.g.~\cite{ste,cai,koy,rpm9,rpm10} and references therein) 
 where they have been christened as the ``phantom'' and/or ``ghost'' fields. These fields (with negative kinetic terms)
automatically lead to the existence of the {\it negative} pressure which happens to be one of the  key characteristic 
features of dark energy (see, e.g.~\cite{zhu,aha} and references therein). Such  kinds of ``exotic'' fields 
have been invoked in the {\it above} cosmological models to explain  the current experimental observations of the accelerated 
expansion of the Universe (see, e.g.~\cite{bps,ast,gon,alba,teg} and references therein for details).

We would like to shed some light on the theoretical contents of our present research work and its key 
{\it differences} with our recent publication~\cite{rpm8}. In our {\it latter} research work, we have 
concentrated on the constraint analysis of the {\it classical} $3D$ Lagrangian density of our theory.
We have devoted time on the {\it proper} gauge-fixing terms and existence of the discrete duality symmetry 
transformations for the gauge-fixed Lagrangian densities. In addition, we have discussed {\it only} the 
off-shell nilpotent (co-)BRST symmetry transformations and a bosonic symmetry transformation that emerges 
out from the anticommutator of {\it these} off-shell nilpotent symmetry transformations. In other words, we have
focused only on the (co-)BRST invariant Lagrangian density~\eqref{1} of our present endeavor where we have 
{\it not} discussed anything about the anti-BRST and anti-co-BRST symmetry transformations. The emphasis 
in~\cite{rpm8} has been laid on the existence and importance of the discrete duality symmetry 
transformations at the classical as well as at the quantum level because we were {\it unaware} of such 
kinds of symmetry transformations in our earlier works (see, e.g.~\cite{rpm6,rpm7}). 
To be precise, our study in~\cite{rpm8} has been very concise in the sense
that {\it all} the symmetry properties of our $3D$ BRST-quantized field-theoretic model have 
{\it not} been discussed unlike our present endeavor where {\it all} the symmetry properties of the 
coupled (but equivalent) Lagrangian densities [cf. Eqs.~\eqref{1},~\eqref{10}]
have been given utmost importance.

Due to the presence of a couple of  {\it non-trivial} CF-type restrictions in~\eqref{22} on our $3D$ theory, we know that the 
{\it Noether} (anti-)BRST and (anti-)co-BRST charges, corresponding to the off-shell nilpotent (anti-)BRST 
and (anti-)co-BRST symmetry transformations, will turn out to be non-nilpotent (see, e.g.~\cite{rpm11} for
details). It will be a nice future endeavor for us to derive the off-shell nilpotent  {\it versions} of the  
(anti-)BRST and (anti-)co-BRST charges from the non-nilpotent versions of the Noether  (anti-)BRST and (anti-)co-BRST 
charges and derive their extended BRST algebra with the {\it other} Noether conserved charges, corresponding to the 
{\it other} infinitesimal and continuous symmetry transformations, of our $3D$ BRST-quantized field-theoretic model. 
This exercise will lead to the physical realizations of the cohomological operators in the language of the  
{\it appropriate} conserved charges. The discussion of the physicality criteria w.r.t the  nilpotent versions 
of the (anti-)BRST and (anti-)co-BRST charges for our $3D$ BRST-quantized model of Hodge theory is yet another 
theoretical direction which we would like to pursue in  our future investigation(s). In  this connection, it 
is worthwhile to lay emphasis on the fact that unlike our earlier works (see, e.g.~\cite{rpm1,rpm2,rpm3,rpm4,rpm5}) 
on the even dimensional (i.e. $D = 2, 4, 6 $) field-theoretic models of Hodge theory, our present system is an 
{\it odd} dimensional (i.e. $ D = 3$) field-theoretic model of Hodge theory. As far as the one $(0 + 1)$-dimensional
($1D$) quantum mechanical (QM) physically interesting systems are concerned, we have been able to prove a couple 
of $1D$ toy models~\cite{rpm12,rpm13} along with a set of interesting $\mathcal {N} = 2$ supersymmetric QM  models 
(see, e.g.~\cite{rpm14,rpm15} and references therein) to be the tractable examples for Hodge 
theory where there is a convergence of ideas from the physics of QM/SUSY-QM and the mathematics of cohomological operators.

\section*{Acknowledgment}

One of us (RPM) would like to thank E. Harikumar, A. K. Rao , S. K. Panja,  Bhagya R. and H. Sreekumar  for useful
discussions on the subject matter of our present investigation.


\appendix

\section{On the Derivation of Eq.~\eqref{25}: Alternative Method}
\renewcommand{\theequation}{A.\arabic{equation}}
\setcounter{equation}{0}
\label{secA}
The CF-type restrictions~\eqref{22} are the essential {\it physical} restrictions on our BRST-quantized $3D$ field-theoretic 
model which are responsible for (i) the absolute anticommutativity 
(i.e. $\{ s_b, \, s_{ab} \} = 0, \, \{ s_d, \, s_{ad} \} = 0$) of the nilpotent (anti-)BRST and (anti-)co-BRST 
symmetry transformations [cf. Eqs.~\eqref{31},~\eqref{32}], and (ii) the existence of the {\it coupled}  Lagrangian 
densities~\eqref{1} and~\eqref{10}. Hence, they should be (anti-)BRST and (anti-)co-BRST invariant for a field-theoretic 
example for Hodge theory. By demanding {\it this} requirement [cf. Eq.~\eqref{24}], we have been able to 
derive the nilpotent (anti-)BRST and (anti-)co-BRST symmetry transformations  for the Nakanishi-Lautrup auxiliary 
fields in~\eqref{25}. The purpose of {\it this} Appendix is to establish that there is an alternative theoretical method 
to derive the {\it exact} transformations that have been listed in~\eqref{25}. In this context, we mention that the nilpotent 
transformations $s_b \bar B_\mu  = - \partial_\mu \lambda, \;  s_{ab} B_\mu = - \partial_\mu \rho$ and 
$s_d \bar {\cal B}_\mu = -\partial_\mu \rho, \; s_ {ad} B_\mu  = - \partial_\mu \lambda$ 
can be derived from the  requirements of the anticommutativity property  between the (anti-)BRST and (anti-)co-BRST
transformations,  respectively. For instance, it can checked that: 
\begin{eqnarray}\label{A.1}
&& \big(s_b\,s_{ab} + s_{ab}\, s_b\big)\,C_\mu = 0 \;\; \Rightarrow \;\; s_b \bar B_\mu = - \,\partial_\mu \lambda, \nonumber\\
&&\big(s_b\,s_{ab} + s_{ab}\, s_b\big)\,\bar C_\mu = 0 \;\;  \Rightarrow \; \;  s_{ab} B_\mu = - \, \partial_\mu \rho, \nonumber\\
&& \big(s_d\,s_{ad} + s_{ad}\, s_d\big) \,C_\mu = 0  \; \; \Rightarrow \; \; s_{ad} {\cal B}_\mu = - \, \partial_\mu \lambda, \nonumber\\
&& \big(s_d\,s_{ad} + s_{ad}\, s_d\big)\,\bar C_\mu = 0 \;\;  \Rightarrow  \;\; s_d  \bar {\cal B}_\mu = -\, \partial_\mu \rho. 
\end{eqnarray}
Similarly, the other fermionic (i.e. nilpotent) symmetry 
transformations in equation~\eqref{25} (i.e.  $s_b \bar {\cal B}_\mu  = 0,\; s_ {ab} {\cal B}_\mu  = 0, \;
s_d \bar B_\mu  = 0,\; s_ {ad} B_\mu  = 0$) can be derived from the requirements of (i) the (anti-)BRST invariance
 of the {\it total} kinetic terms for the Abelian 1-form gauge field $A_\mu$ [cf. Lagrangian densities~\eqref{1} and~\eqref{10}],  and 
(ii) the (anti-) co-BRST invariance of total gauge-fixing terms for the Abelian 2-form gauge field $B_{\mu\nu}$ 
[cf. Lagrangian densities~\eqref{1} and~\eqref{10}], respectively. These  requirements can be mathematically 
expressed as: 
\begin{eqnarray}\label{A.2}
&& s_b \bigg[\dfrac{1}{2}\, \bar {\cal B}^\mu \bar {\cal B}_\mu + \bar {\cal B}^\mu \Big(\varepsilon_{\mu\nu\sigma} \,\partial^\nu A^\sigma 
+ \dfrac{1}{2}\, \partial_\mu \widetilde \phi \Big) \bigg] = 0 \;\; \quad \Rightarrow\;\; \quad s_b \bar {\cal B}_\mu = 0, \nonumber\\
&& s_{ab} \bigg[\dfrac{1}{2}\, {\cal B}^\mu {\cal B}_\mu 
- {\cal B}^\mu \Big(\varepsilon_{\mu\nu\sigma} \,\partial^\nu A^\sigma 
- \dfrac{1}{2}\, \partial_\mu \widetilde \phi \Big) \bigg] = 0 \; \quad \Rightarrow\; \quad s_{ab} {\cal B}_\mu = 0, \nonumber\\
&& s_d \bigg[- \dfrac{1}{2}\, \bar B^\mu \bar B_\mu - \bar B^\mu \Big( \partial^\nu B_{\nu\mu} 
+ \dfrac{1}{2}\, \partial_\mu \phi \Big) \bigg] = 0  \; \; \quad \Rightarrow\; \quad s_d \bar B_\mu = 0, \nonumber\\
&& s_{ad} \bigg[- \dfrac{1}{2}\, B^\mu B_\mu +  B^\mu \Big( \partial^\nu B_{\nu\mu} 
- \dfrac{1}{2}\, \partial_\mu \phi \Big)  \bigg] = 0 \; \quad \Rightarrow\; \quad  s_{ad} B_\mu = 0.
\end{eqnarray}
It is obvious, from the equations~\eqref{A.1} and \eqref{A.2}, that we have derived precisely {\it all} the 
off-shell nilpotent transformations that have been listed in~\eqref{25}.

We end this Appendix with the following remarks. First of all, the requirement  of the absolute anticommutativity 
between (i) the nilpotent BRST and anti-BRST symmetry transformations, and (ii) the co-BRST and anti-co-BRST 
symmetry transformations, is one of the sacrosanct properties of the BRST formalism. Physically, the above requirements 
imply the linear independence of the (anti-)BRST and (anti-)co-BRST symmetry transformations at the quantum level. 
Hence, our results in the equation~\eqref{A.1} are correct. Second, the {\it gauge-invariant} kinetic terms for the Abelian 
1-form and 2-form gauge fields owe their origin to the exterior derivative of differential geometry. Furthermore, the {\it classical} 
gauge transformations of our theory are elevated to the (anti-)BRST transformations at the quantum level within 
the framework of BRST formalism. Hence, the kinetic terms for {\it both} the gauge fields must be invariant under the (anti-)BRST 
symmetry transformations at the quantum level. This is what we have taken into account in the {\it first} two entries 
of our equation~\eqref{A.2}. Finally, the gauge-fixing terms of the Abelian 1-form and 2-form gauge fields are generated 
by the application of the co-exterior derivative of differential geometry on the above gauge fields. Hence, they should 
remain invariant under the (anti-)co-BRST symmetry transformations at the quantum level. We have taken into 
account this fact in the {\it last} two entries of our equation~\eqref{A.2}. We, ultimately, conclude that there are two different 
ways to derive the off-shell nilpotent symmetry transformations~\eqref{25}.


\section{On Connections Between $s_{(a)d}$ and $s_{(a)b}$: Direct Application of Discrete Symmetry Transformations}
\label{secB}
\renewcommand{\theequation}{B.\arabic{equation}}
\setcounter{equation}{0}
In this Appendix, we establish the connections between the off-shell nilpotent (i.e. $s_d^2 = 0, \, s_b^2 = 0 $) 
(co-)BRST symmetry transformations [cf. Eqs.~\eqref{2},~\eqref{3}] by exploiting the {\it direct} application of 
the discrete duality symmetry transformations  for the Lagrangian density~\eqref{1} along with the transformations: 
$*\, s_b = s_d, \; *\, s_d = -\, s_b $ where $*$ stands for the discrete duality symmetry transformations 
[cf. Eq.~\eqref{6}]. To corroborate {\it this} statement, let us take a simple example: $s_b A_\mu = \partial_\mu C$ 
[cf. Eq.~\eqref{3}]. Applying the discrete transformations~\eqref{6} on this relationship, we obtain the following
explicit relationship, namely;
\begin{eqnarray}\label{B.1}
* \big (s_b A_\mu \big )= * \big (\partial_\mu C\big ) \;\; \Rightarrow \;\;
 \big ( *s_b \big) \big (A_\mu \big ) = \partial_\mu \big (* C \big),
\end{eqnarray}
where the discrete duality symmetry transformations~\eqref{6}, corresponding to the $*$ operation, act only on the {\it internal}
symmetry transformation operator $s_b$ and the fields of the Lagrangian density~\eqref{1}. These transformations do {\it not} act on the 
spacetime derivative operator (i.e. $\partial_\mu$) because our $3D$ flat spacetime manifold stays in the background and it
does {\it not} participate in our present discussions on the symmetry properties of our $3D$ field-theoretic model. Taking into
account: $ *\, s_b = s_d$ and the discrete duality symmetry transformations~\eqref{6}, we observe that the following is {\it true}, namely; 
\begin{eqnarray}\label{B.2}
&& s_d  \Big(\mp \frac{i}{2} \,\varepsilon_{\mu\nu\sigma}\,  B^{\nu\sigma} \Big) \,= \,\partial_\mu \big(\mp \, i\, \bar C\big) 
 \;\;\Rightarrow\;\; s_d B_{\mu\nu} = \varepsilon_{\mu\nu\sigma}\,\partial^\sigma \bar C.
\end{eqnarray}
Hence, it is crystal clear that we have obtained the co-BRST symmetry transformation for the Abelian 
2-form antisymmetric tensor gauge field $B_{\mu\nu}$ from the BRST symmetry transformation for the 
Abelian 1-form vector gauge field $A_\mu$. In exactly similar fashion, if we start with the co-BRST 
symmetry transformation: $s_d B_{\mu\nu} = \varepsilon_{\mu\nu\sigma} \,\partial^\sigma \bar C $ and 
apply the transformations: $*\, s_d = -\, s_b $ and~\eqref{6} on {\it this} relationship, we obtain the following 
\begin{eqnarray}\label{B.3}
&&\big(*\, s_d \big) \, \big(*\, B_{\mu\nu} \big)= \varepsilon_{\mu\nu\sigma} \partial^\sigma \big (*\, \bar C \big) 
\;\;\Rightarrow \;\; -\, s_b \big(\pm i\, \varepsilon_{\mu\nu\sigma} \, A^\sigma \big) 
= \varepsilon_{\mu\nu\sigma} \, \partial^\sigma \, \big(\mp\, i\, C \big), \quad
\end{eqnarray}
which, ultimately, leads to the derivation of the off-shell nilpotent BRST symmetry transformation: 
$s_b A_\mu = \partial_\mu C$ for the vector field $A_\mu$. Thus, we have established a very beautiful relationship: 
$s_b A_\mu = \partial_\mu C \; \Leftrightarrow \; s_d B_{\mu\nu} = \varepsilon_{\mu\nu\sigma}\, \partial^\sigma \bar C$ 
between the BRST symmetry transformation for the Abelian 1-form vector field $A_\mu$ and the co-BRST 
symmetry transformation for the Abelian 2-from antisymmetric field $B_{\mu\nu}$. This procedure can be 
repeated in a straightforward manner and we obtain the following complete set of relationships between 
the BRST and co-BRST symmetry transformations [cf. Eqs.~\eqref{3},~\eqref{2}], namely;
\begin{eqnarray}\label{B.4}
&& s_b A_\mu = \partial_\mu C \; \Leftrightarrow \;
 s_d B_{\mu\nu} = \varepsilon_{\mu\nu\sigma}\, \partial^\sigma \bar C,  \qquad
 s_b C_\mu = -\, \partial_\mu \beta \; \Leftrightarrow \; s_d \bar C_\mu = -\, \partial_\mu \bar \beta, \nonumber\\
&& s_b B_{\mu\nu} = - \big( \partial_\mu C_\nu -  \partial_\nu C_\mu \big ) \,
\Leftrightarrow \, s_d A_\mu = - \varepsilon_{\mu\nu\sigma}\, 
\partial^\nu \bar C^\sigma,  \qquad
 s_b \bar C_\mu = B_\mu  \, \Leftrightarrow \, s_d  C_\mu = -\, {\cal B}_\mu, \nonumber\\
&&s_b \bar \beta = -\, \rho \; \Leftrightarrow \; s_d  \beta = -\, \lambda, \qquad
 s_b \phi = \lambda \; \Leftrightarrow \; s_d \widetilde \phi = \rho, 
\qquad s_b \bar C = B \; \Leftrightarrow \; s_d C = {\cal B}, \nonumber\\
&& s_b \big [B_\mu, \, {\cal B}_\mu, \, B, \, {\cal B}, \, \widetilde \phi, \, \beta, \, C, \, \rho, \, \lambda \big ] = 0 
\;\Leftrightarrow \; s_d \big [{\cal B}_\mu, \, B_\mu, \, {\cal B}, \, B,  \,  
\phi, \, \bar \beta, \, \bar C, \,  \lambda, \, \rho \big ] = 0. \;\;
\end{eqnarray}
Thus, ultimately, we have been able to establish a connection between the off-shell nilpotent BRST and co-BRST symmetry 
transformation operators of the Lagrangian density~\eqref{1} by exploiting the discrete transformations~\eqref{6} 
and  taking into account the crucial inputs: $*\, s_b = s_d, \; *\, s_d = -\, s_b $.

To complete our discussion, we now focus on establishing a  connection between the off-shell nilpotent 
(i.e. $s_{ab}^2 = 0, \; s_{ad}^2 = 0 $) anti-BRST ($s_{ab}$) and anti-co-BRST ($s_{ad}$) symmetry transformation 
operators [cf. Eqs.~\eqref{11},~\eqref{12}] by using: $*\, s_{ab} = s_{ad}, \; *\, s_{ad} = -\, s_{ab} $ and the 
analogues of the discrete duality symmetry transformations~\eqref{6} for the Lagrangian density~\eqref{10}. Adopting 
exactly the same kind of procedure as we have chosen for establishing the relationship [cf. Eq.~\eqref{B.4}] 
between the (co-)BRST symmetry transformation operators, we obtain the following complete relationships between
$s_{ab}$ and  $s_{ad}$, namely;
\begin{eqnarray}\label{B.5}
&& s_{ab} A_\mu = \partial_\mu \bar C \; \Leftrightarrow \;
s_{ad} B_{\mu\nu} = \varepsilon_{\mu\nu\sigma}\, \partial^\sigma  C, \qquad
s_{ab} \bar C_\mu = -\, \partial_\mu \bar \beta \; \Leftrightarrow \; s_{ad}  C_\mu  =  \partial_\mu \beta,  \nonumber\\
&& s_{ab} B_{\mu\nu} = - \big( \partial_\mu \bar C_\nu -  \partial_\nu \bar C_\mu \big ) 
\Leftrightarrow  s_{ad} A_\mu = - \varepsilon_{\mu\nu\sigma}\, \partial^\nu  C^\sigma,  \qquad
s_{ab}  C_\mu  = \bar B_\mu  \Leftrightarrow  s_{ad} \bar  C_\mu  = - \bar {\cal B}_\mu, \nonumber\\
&&s_{ab}  \beta = - \lambda \; \Leftrightarrow \; s_{ad} \bar \beta = \rho, \qquad 
s_{ab} \phi  = \rho \;  \Leftrightarrow  \;  s_{ad} \widetilde \phi  = \lambda, \qquad
 s_{ab}  C = - B \; \Leftrightarrow \; s_{ad} \bar C =  {\cal B}, \nonumber\\
&& s_{ab} \big [B_\mu, \, {\cal B}_\mu, \, B, \, {\cal B}, \, \widetilde \phi, \, \bar \beta, \,\bar C, \, \rho, \, \lambda \big ] = 0 
\;\Leftrightarrow \, s_{ad} \big [{\cal B}_\mu, \, B_\mu, \, {\cal B}, \,  B, \,   \phi, \,  \beta, \,  C, \, \lambda, \, \rho \big ] = 0. 
\end{eqnarray}
Thus, finally, we have been able to establish an intimate connection between the off-shell nilpotent anti-BRST and anti-co-BRST 
transformation operators [cf. Eqs.~\eqref{11},~\eqref{12}] by exploiting the {\it direct} applications 
of the discrete duality  transformation operator for the Lagrangian density~\eqref{10} which are the analogues 
of~\eqref{6} along with the crucial inputs: $*\, s_{ab} = s_{ad}, \; *\, s_{ad} = -\, s_{ab} $.

We wrap-up this Appendix with the following clinching remarks. First, the forms of the duality symmetry 
transformations: $*\, s_b = s_d, \; *\, s_d = -\, s_b $ and  $*\, s_{ab} = s_{ad}, \; *\, s_{ad} = -\, s_{ab} $ 
are just like the duality symmetry transformations of the source-free $4D$ Maxwell's equations where the
 electric field (${\bf E}$) and magnetic field  (${\bf B}$) obey the relationships: 
${\bf E} \to {\bf B}, \; {\bf B} \to -\, {\bf E}$. Second, we  have obtained the alternative relationships between 
(i) the BRST and co-BRST symmetry transformation operators  in Section~\ref{sec2} [cf. Eqs.~\eqref{7},~\eqref{9}], and 
(ii) the anti-BRST  and anti-co-BRST symmetry transformations in Section~\ref{sec3} [cf. Eq.~\eqref{15}] which are 
totally {\it different} from the contents of our present Appendix. Third, the choices of the 
 duality symmetry transformations on the off-shell nilpotent transformation operators 
(e.g. $*\, s_b = s_d, \; *\, s_d = -\, s_b $ and  $*\, s_{ab} = s_{ad}, \; *\, s_{ad} = -\, s_{ab} $) 
have been made by exploiting the theoretical strength of the trial-and-error method. There are {\it no}
basic principles of any kinds that are involved and/or invoked in these derivations. Fourth, in our 
equations~\eqref{B.4} and~\eqref{B.5}, we have derived the connections amongst the nilpotent symmetry 
transformations~\eqref{2},~\eqref{3},~\eqref{11} and~\eqref{12} {\it only}. Such type of relationships are true
for the off-shell nilpotent symmetry transformations~\eqref{25}, too. For instance, we observe that the following
\begin{eqnarray}\label{B.6}
&& s_{ab} {\cal B}_\mu = 0 \;\; \Leftrightarrow \;\;
 s_{ad} B_{\mu} = 0,  \qquad
 s_{ab} B_\mu = -\, \partial_\mu \rho \;\; \Leftrightarrow \;\; s_{ad}  {\cal B}_\mu  =  - \,\partial_\mu \lambda,  \nonumber\\
&& s_{b} \bar {\cal B}_\mu = 0 \;\; \Leftrightarrow \;\;
 s_{d} \bar B_{\mu} = 0,  \qquad
s_{b} \bar B_\mu = -\, \partial_\mu \lambda \;\;\Leftrightarrow \;\; s_{d}  \bar {\cal B}_\mu  =  - \,\partial_\mu \rho,
\end{eqnarray}
are {\it also} true due to the {\it direct} application of the discrete symmetry transformations~\eqref{6} 
for the Lagrangian density~\eqref{1}  and their analogues for the Lagrangian density~\eqref{10} along with the inputs: 
$*\, s_b = s_d, \; *\, s_d = -\, s_b, \;*\, s_{ab} = s_{ad}, \; *\, s_{ad} = -\, s_{ab} $. 
Finally, we point out that, in our previous work~\cite{rpm8}, we have discussed {\it concisely} such kind of 
{\it direct} relationship between the off-shell nilpotent BRST and co-BRST symmetry transformation operators. However, 
these discussions are {\it not} as elaborate as what we have done in our present endeavor where the full set 
of the off-shell nilpotent (anti-)BRST as well as the (anti-)co-BRST symmetry operators have been taken into 
account (cf. Sections~\ref{sec2},~\ref{sec3},~\ref{sec7},~\ref{sec8}).



\let\doi\relax

\end{document}